%% file: main.tex
\documentclass[sigconf,nonacm]{acmart}

\setcopyright{none}

\usepackage{tabularray}
\usepackage{tikz}
\usepackage{amsmath}
\usepackage{graphicx}
\usepackage{xcolor}
\usepackage{color, soul}
\usepackage{multirow}
\usepackage{multicol}
\usepackage{subfig}
\usepackage{tabularx}
\usepackage{hyperref}
\hypersetup{hidelinks}
\usepackage{xspace}
\usepackage{appendix}
\usepackage{stfloats}
\usepackage{hhline}
\usepackage{capt-of}
\usepackage{xcolor}

\newcommand*\circled[1]{\tikz[baseline=(char.base)]{
            \node[shape=circle,draw,inner sep=0.1pt] (char) {#1};}}

\newcommand{\ourtool}{\textit{Heimdallr}\xspace}

\begin{document}

\title{Heimdallr: Fingerprinting SD-WAN Control-Plane Architecture via Encrypted Control Traffic}
\thanks{\textit{This is the author’s accepted manuscript of the article published in 
the Proceedings of the 38th Annual Computer Security Applications Conference (ACSAC '22), 
DOI: \url{https://doi.org/10.1145/3564625.3564642}.}}

\author{Minjae Seo$^1$, Jaehan Kim$^2$, Eduard Marin$^3$, Myoungsung You$^2$, Taejune Park$^4$, Seungsoo Lee$^5$, Seungwon Shin$^2$, and Jinwoo Kim$^6$}
\affiliation{\vspace{0.1cm}
\institution{$^1$The Affiliated Institute of ETRI, $^2$KAIST, $^3$Telefonica Research, $^4$Chonnam National University,\\$^5$Incheon National University, $^6$Kwangwoon University}
\country{} 
}

\authornote{Jinwoo Kim is the corresponding author.}

\renewcommand{\shortauthors}{Seo et al.}

\input{0_abstract}


\maketitle

\input{1_introduction}

\input{2_motiv_background}

\input{3_design}
\input{4_eval}

\input{5_use_case}

\input{6_related_work}

\input{7_conclusion}

\bibliographystyle{ACM-Reference-Format}
\bibliography{references}

\input{8_appendix}

\end{document}

%% file: 0_abstract.tex
\begin{abstract}

Software-defined wide area network (SD-WAN) has emerged as a new paradigm for steering a large-scale network flexibly by adopting distributed software-defined network (SDN) controllers. The key to building a logically centralized but physically distributed control-plane is running diverse cluster management protocols to achieve consistency through an exchange of control traffic. Meanwhile, we observe that the control traffic exposes unique time-series patterns and directional relationships due to the operational structure even though the traffic is encrypted, and this pattern can disclose confidential information such as control-plane topology and protocol dependencies, which can be exploited for severe attacks. With this insight, we propose a new SD-WAN fingerprinting system, called \textit{Heimdallr}. It analyzes periodical and operational patterns of SD-WAN cluster management protocols and the context of flow directions from the collected control traffic utilizing a deep learning-based approach, so that it can classify the cluster management protocols automatically from miscellaneous control traffic datasets. Our evaluation, which is performed in a realistic SD-WAN environment consisting of geographically distant three campus networks and one enterprise network shows that \textit{Heimdallr} can classify SD-WAN control traffic with $\geq$ 93\%, identify individual protocols with $\geq$ 80\% macro F-1 scores, and finally can infer control-plane topology with $\geq$ 70\% similarity.

\end{abstract}

%% file: 1_introduction.tex
\section{Introduction}

Today, many large enterprises, telecoms, and cloud providers operate dedicated wide area networks (WANs) to connect their data centers and remote sites distributed across the globe. Microsoft, Google, and Facebook are prominent examples of companies that constructed their own (private) WANs~\cite{jain2013b4,greenberg2015sdn,ferguson2021orion}. These WANs typically connect tens to hundreds of locations, have global spans, and provide low-latency, high throughput inter-data center communication. Due to the increasing size and complexity of such networks, network operators have begun to embrace Software-Defined Networking (SDN) when building their WANs~\cite{jain2013b4, hong2018b4, hong2013achieving}. Unlike traditional WANs, SD-WANs simplify the process of building and managing connections between sites, and provide more flexibility, greater programmability, centralized control as well as improved monitoring while also lowering operational costs.

As different sites within an SD-WAN exchange a large amount of control and data traffic (much of which is sensitive), WANs are known to be very attractive targets for adversaries to carry out attacks. Indeed, recent disclosures have shown that governmental agencies (and similar entities) were able to eavesdrop SD-WAN traffic by tapping on intercontinental fiber links between multiple sites~\cite{wiretapping}. Major companies like Google have recently recognized the severity of this threat (not only in the links that interconnect their sites but also in information exchanges that occur inside their sites), and have started to use protocols like MACsec and IPSec in order to encrypt SD-WAN traffic not just at the application layer, but also at the network and link layers~\cite{aws_macsec}.

While encrypting SD-WAN traffic is an effective way to alleviate such attacks, encryption alone is \emph{not} sufficient as traffic analysis attacks over encrypted data are still possible. Crucially, any information that is preserved when traffic is encrypted, such as traffic volume, packet sizes, and packet timings, can disclose many insights about the network traffic. In the context of an SD-WAN, adversaries can analyze the (encrypted) traffic patterns in an attempt to obtain previously unknown information about the SD-WAN such as its control-plane topology or the cluster management protocols being used. These are both regarded as sensitive information, which, in the hands of adversaries, can be used to find valuable targets or to discover the presence of a protocol which is known to contain some vulnerabilities.

To the best of our knowledge, we are the first to show that the control-plane topology of an SD-WAN and the cluster management protocols it uses, can be inferred by analyzing the (encrypted) traffic exchanged between sites. In this paper, we propose \ourtool{}, a deep-learning-based system suitable to fingerprint SD-WANs, leveraging the fact that SD-WAN control traffic maintains perceptible time-series patterns due to the need for periodic synchronization and directional relationships among the distributed controllers. Our deep learning-based classification models are based on a Long Short-Term Memory (LSTM)~\cite{hochreiter1997long}. Our experiments show that LSTM-based models are highly effective for reflecting time-series patterns and directional relationships of the SD-WAN control traffic comparing with other neural networks and traditional time-series modeling algorithms.

We implement a full prototype of \ourtool{} and conduct extensive experiments in a real SD-WAN testbed consisting of commercial SDN-enabled switches and popular SDN controllers~\cite{onos_web}. The testbed is composed of four different sites (three campus networks and one enterprise network) with an average distance of 170 kilometers. We have accumulated about 57,300,000 packets generated from several SDN applications while deploying the cluster management protocols for SD-WAN control traffic and using CAIDA traces along with various enterprise services/consensus protocols to represent the data traffic. We show \ourtool{}'s effectiveness by performing an extensive set of experiments in three different testing environments. We also demonstrate that \ourtool{} maintains robust performance even when several defense systems are used. From this, we could verify realistic scenarios where adversaries collect only parts of the traffic and demonstrate the feasibility of our fingerprinting system.

\noindent\textbf{Contributions:} Our main contributions are summarized as follows:
\begin{itemize}
    \item The in-depth exploration of a new attack vector in SD-WAN, which allows adversaries to infer control-plane topology and protocol dependencies solely from encrypted control traffic.

    \item The proposal of a deep learning-based SD-WAN fingerprinting system, \ourtool{}, which includes noticeable experiment results with several defense systems performed in a realistic SD-WAN.

    \item The extensive experiments in a hardware-based private SD-WAN testbed utilizing the developed fingerprinting system to demonstrate the effectiveness of the posed vulnerabilities. 
\end{itemize}

%% file: 2_motiv_background.tex
\section{Background and Motivation}
\label{sec:background}

In this section, we briefly introduce the necessary background of SD-WAN and cluster management protocols.

\begin{figure}[t]
    \centering
    \includegraphics[width=0.85\linewidth]{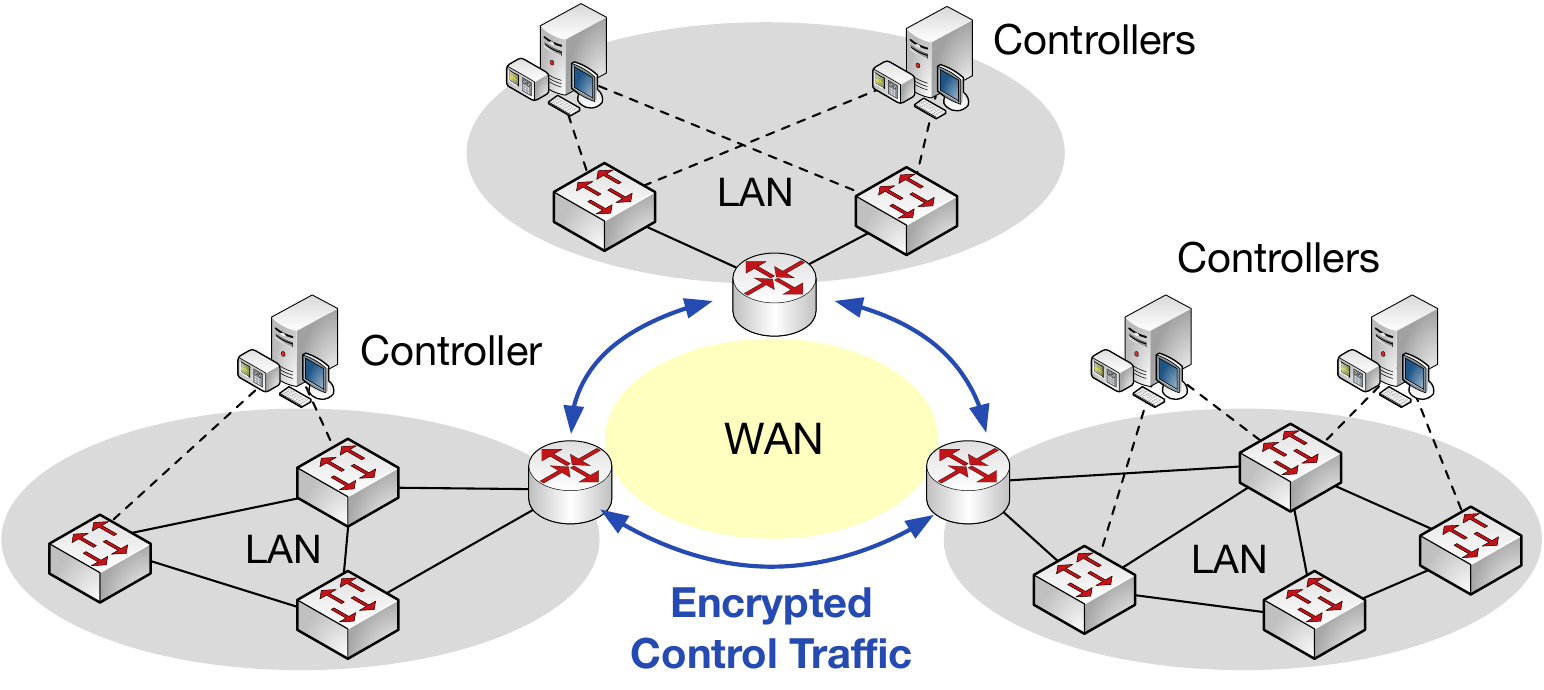}
    \caption{Distributed SDN controllers for SD-WAN.}
    \label{fig:sdwan_model_a}
    \vspace{-0.2in}
\end{figure}

\subsection{SD-WAN}
\label{ss:sdwan}

Over the last decade, Software Defined Networking (SDN) has gained popularity both in academia and industry, causing network operators to use it in their data center~\cite{tammana2015cherrypick}, telco~\cite{cord}, enterprise networks~\cite{casado2007ethane}. The SDN paradigm separates the network's intelligence (i.e., control plane) from the forwarding functionality of networking devices (i.e., data plane). By doing so, it allows to place the network's intelligence into a logically centralized controller whose functionalities can be extended via applications using a set of standard APIs---widely known as Northbound and Southbound interfaces. Thanks to this innovation, network administrators can directly control network devices across the entire network, simplifying the management task.

More recently, the SDN paradigm has been adopted in WAN to achieve application-aware traffic engineering in a geographically large area~\cite{jain2013b4, hong2013achieving}. As networks grow and sites are located in long-distance areas, a single SDN controller faces coverage and scalability issues~\cite{yang2019software}. To address this problem, administrators have adopted \emph{distributed} controllers to operate SD-WAN efficiently and safely. Figure~\ref{fig:sdwan_model_a} illustrates the concept of SD-WAN with the controllers deployed in multiple sites. Notice that the controllers exchange encrypted control traffic over WAN using \emph{East and West interfaces} to support controller-to-controller communication.

\begin{figure}[t]
    \centering
    \includegraphics[width=0.90\linewidth]{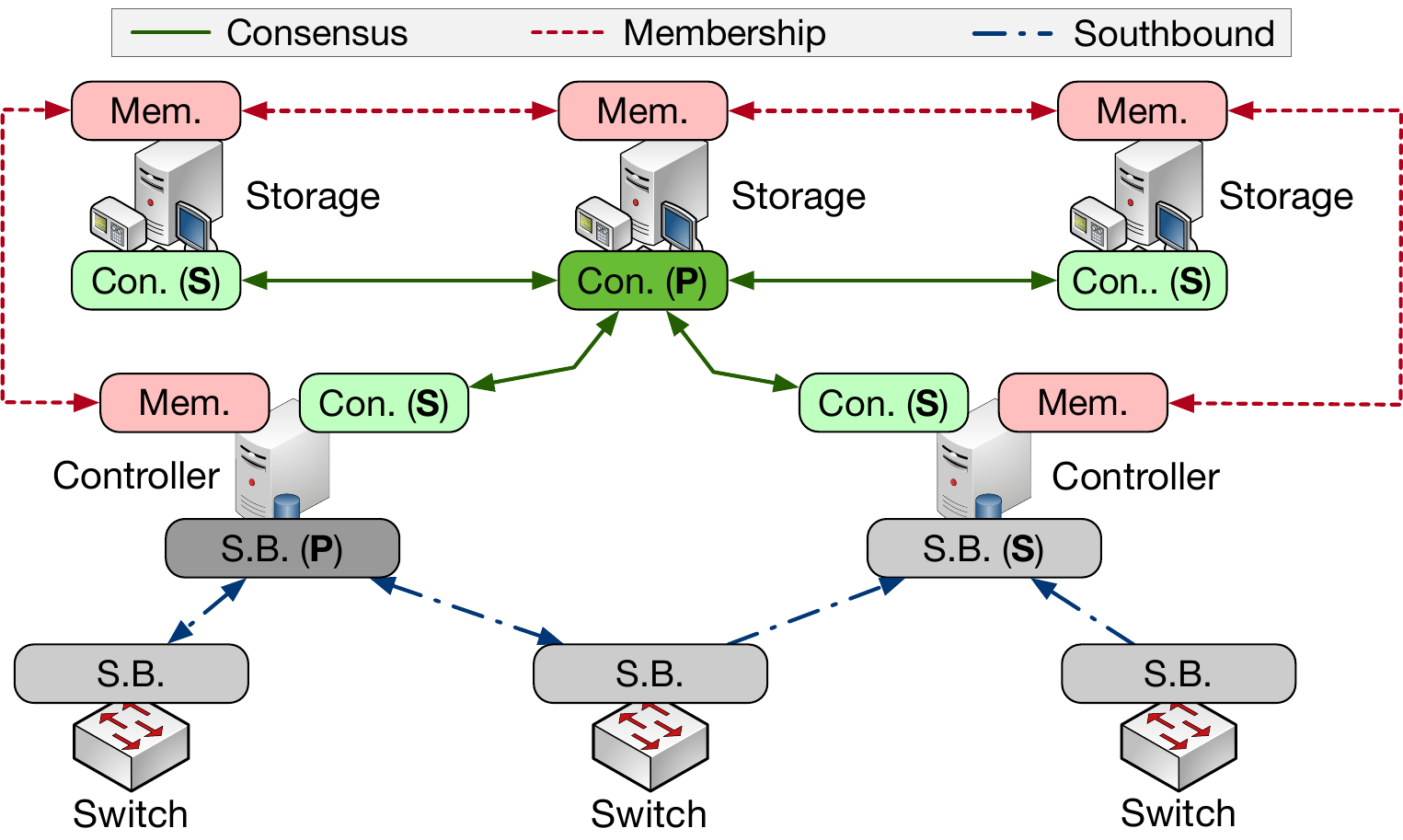}
    
    \footnotesize{$\ast$ \textbf{P}: Primary, \textbf{S}: Secondary}
    \caption{An example of control-plane topology in SD-WAN.}
    \label{fig:protocol_dependency}
    \vspace{-0.2in}
\end{figure}

\subsection{Cluster Management Protocols}
\label{ss:cluster}

Most popular distributed controllers employ a hierarchical architecture to achieve better flexibility and scalability in the cluster\footnote{In this paper, we define a \emph{cluster} as a set of nodes in the control plane.}~\cite{jain2013b4, ferguson2021orion, onos_web, odl_web}. Typically we can distinguish between two types of nodes; (i) \emph{storage nodes} used to synchronize states between all nodes; and (ii) \emph{controller nodes} that interact with the switches and are used to enforce network policies. These nodes collectively run a series of cluster management protocols that are crucial for the correct functioning of the SD-WAN (see Figure~\ref{fig:protocol_dependency}), including (i) a membership protocol, (ii) a consensus protocol, and (iii) a southbound protocol.

The \textbf{membership protocol} is responsible for periodically checking the status of cluster nodes to see if they are still alive. This information is fundamental to maintaining an up-to-date list of currently available nodes. For this purpose, nodes exchange probe messages with each other periodically, which can be achieved by using either (i) broadcast or (ii) unicast. For example, the ONOS SDN controller resorts to the SWIM~\cite{das2002swim} protocol, which is based on unicast probing one by one all nodes. On the contrary, OpenDaylight utilizes the heartbeat mechanism of the Raft protocol, which performs broadcast transmission from one node to the others.

The \textbf{consensus protocol} manages state synchronization to keep consistency among storage nodes. So far, diverse consensus protocols have been employed, such as Raft~\cite{onos_web,odl_web}, Paxos~\cite{jain2013b4}, Pub/Sub, and Zab~\cite{onos_web}. The common pattern observed from those protocols is that one node (e.g., publisher, leader) produces a message containing a current state while other nodes (e.g., subscriber, follower) consume it. Then, the other nodes update the states and reply with acknowledgment when receiving the message.

The \textbf{southbound protocol} enables communications between controllers and switches to populate forwarding states and collect network statistics. The most widely used one is OpenFlow (OF)~\cite{openflow_1_3}, the de facto standard in SDN. The distinguishable OF feature suitable to a distributed environment is \emph{mastership}; when a switch is connected, controllers compete with each other to gain a master role that has write permission in the switch. The nodes that fail to gain the mastership become slave roles with read-only permissions.

Overall, we can see that cluster management protocols have prominent features. First, nodes are distinguished into two different roles: \emph{primary} and \emph{secondary}. Note that the former is responsible for communicating with all other nodes while the latter does so only with the primary node. Second, protocols are involved in complicated \emph{dependencies} as each protocol is in charge of interactions between different components. Figure~\ref{fig:protocol_dependency} illustrates an example of roles and dependencies in the cluster.

\subsection{Motivation}
\label{ss:motivation}

If adversaries manage to obtain information about the roles and dependencies within the cluster, they can infer the confidential control-plane topology of an SD-WAN and later on use it to mount several attacks more effectively, efficiently, and stealthily.

First, adversaries can mount target-guided DDoS attacks. Suppose adversaries discover which IP addresses the cluster nodes have, and especially, which IP address a \emph{primary} node has, through the inferred control-plane topology. Here, they can attempt to cut off the links around the primary node using stealthy DDoS attacks~\cite{kang2013crossfire,studer2009coremelt,cao2019crosspath}, or directly terminate the node operation using off-path TCP injection attacks~\cite{feng2020off,gilad2014off}. As the primary node plays an important role in the cluster, simply targeting a single node can break the entire SD-WAN operation (we demonstrate it in Section~\ref{sec:use_cases}).

Second, adversaries can exploit known vulnerabilities of inferred protocols. Existing works suggest that cluster management protocols exhibit abnormal behaviors when the environment changes unexpectedly ~\cite{zhang2017raft,scott2016minimizing}. For example, the Raft protocol elects a leader (i.e., primary node) whose \emph{term} is higher than others. According to Raft specification~\cite{ongaro2014search}, the leader withdraws its current leadership when discovering a node whose term is higher than itself. However, two leaders can be elected at different places if a network is partitioned due to a link failure. As Raft does not allow multiple leaders, they will eventually compete with each other forever, thereby interrupting the entire cluster operation~\cite{zhang2017raft}. Adversaries who learn that Raft is being used in the target SD-WAN can attempt to disconnect a specific link between nodes to trigger this vulnerability.

%% file: 3_design.tex
\section{Fingerprinting SD-WAN}

\subsection{Threat Model}
\label{sec:threatmodel}

We consider a wide area network (WAN) that connects multiple sites (spanned across geographically distant locations) of one enterprise over dedicated, encrypted tunnels which can be created at layer 2 (e.g., using MACsec) or at layer 3 (e.g., using IPsec). Within each SD-WAN site, we assume network operators use in-band control~\cite{in-band}, meaning that some links in the network carry both control and data traffic. Note that in-band control is widely used in large SDN-based networks as it reduces the cost of building a dedicated control network and simplifies network maintenance considerably~\cite{braun2014software, ferguson2021orion, cao2019crosspath, xu2017attacking}. We also assume each site has one or multiple controllers (for availability and scalability reasons) and that adversaries do \emph{not} have any prior knowledge about the target SD-WAN (e.g., the SDN controller being used).

The goal of adversaries is to infer sensitive information about the SD-WAN, such as the control-plane topology or the underlying cluster management protocols, which later on can be used to perform attacks more effectively, efficiently, and stealthily. To achieve their goal, adversaries can capture SD-WAN traffic and then try to get insights about the SD-WAN by analyzing the observed (encrypted) traffic patterns. Note that, because of the cryptographic protection, adversaries are unable to inject or modify packets. Adversaries can be present within an SD-WAN site (we denote these as on-site adversaries) or can be located in the path between a pair of SD-WAN sites---the latter are denoted as network adversaries. 

We believe both attack scenarios are realistic. \emph{On-site adversaries} can control one or more virtual machines or containers, and use them to launch attacks against switches~\cite{10.1145/3185467.3185468}. In recent years, many serious vulnerabilities have been found in SDN switches~\cite{10.1145/3386367.3431313} (including software switches), also in the context of SD-WANs~\cite{sdn_wan_vuln, vulnswitch1, vulnswitch2}. Another possibility for adversaries to eavesdrop SD-WAN traffic would be to carry out any of the existing topology attacks (in particular link fabrication attacks) against SDN networks~\cite{marin2019depth, hong2015poisoning}. Meanwhile, network adversaries can perform eavesdropping attacks on SD-WAN traffic either passively (if they are in one of the ASes in the path the SD-WAN traffic traverses) or actively (if they are in an AS that is \emph{not} in the traversed path)\footnote{While not the main type of adversary we consider in this paper, there might also be more powerful in-network adversaries (e.g., governmental agencies and similar entities) who can perform wiretapping attacks directly on intercontinental fiber optics~\cite{wiretapping}.}. In the latter case, network adversaries can leverage BGP hijacking attacks, which are becoming increasingly frequent today~\cite{bgp_hijacking}, in order to redirect the SD-WAN traffic to themselves.

\subsection{Technical Challenges}

\noindent\textbf{Strawman Solution}: Adversaries could simply rely on traditional
rule-based strategies~\cite{finsterbusch2013survey, dainotti2012issues} to classify cluster management protocols based on empirical rules. However, as SD-WAN operators might be able to use custom configurations (e.g., by changing the port numbers), this approach can lead to accuracy deterioration when processing unmatched traffic policies~\cite{mao2017routing, mao2018novel, wang2017end}. Unlike the traditional method, the state-of-the-art deep learning-based system can learn hidden patterns in the control traffic itself, so the aforementioned problems can be effectively handled. In order to take advantage of a deep learning-based system in our work, we need to address the following three challenges:

\label{sec:system_challenge}

\begin{figure*}[t]
    \centering
    \includegraphics[width=0.95\textwidth]{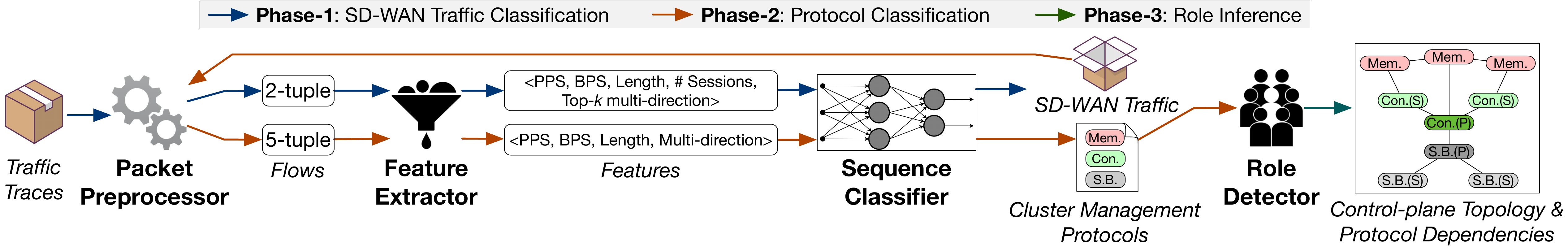}
    \vspace{-0.1in}
    \caption{\ourtool{} system overview and its workflow that consists of three phases: \protect\circled{1} SD-WAN Control Traffic Classification, \protect\circled{2} Protocol Classification, and \protect\circled{3} Role Inference.}
    \label{fig:system_overview}
    \vspace{-0.15in}
\end{figure*}

\noindent\textbf{C1. How to prune data traffic noise: } Nowadays, data traffic is very dynamic and diverse due to the pervasive usage of many different applications and services within networks~\cite{anttila2016youtube, coppola2016connected}. Note that the SD-WAN traffic captured by adversaries contains not only control traffic but also data traffic; it may happen that the data traffic patterns accidentally overlap with the cluster management protocols' patterns. Thus, the classification model may falsely detect the given traffic. In this context, the challenge is how to select the features to be used in the classifier such that false positives and false negatives are minimized considerably.

\noindent\textbf{C2. How to distinguish between cluster management protocols:} Once adversaries classify control and data traffic, they need to identify which cluster management protocols are being used. However,
SD-WAN control packets exchanged between cluster nodes, specifically consensus and membership control traffic, are mixed in a single connection (sharing the same port).
Also, southbound control traffic between a controller and a switch is carried by another single connection. 
Thus, it is impossible to classify application-level protocols using simple network identifiers (e.g., port numbers).

\noindent\textbf{C3. How to determine the role of nodes:} After classifying the cluster management protocols, the question of how to define the primary-secondary roles of nodes in the cluster still remains. To infer the role of each node, we need to design additional deep learning models following the protocol classifier.
However, the number of flows sent by the primary node is significantly lower compared to those sent by the secondary nodes,
and handling sparse labels can make the model fall into overfitting~\cite{guo2013probabilistic}. Thus, models cannot be trained effectively to identify the role of each node.

\subsection{Overview of \ourtool{}}

Based on the given challenges, we devise an advanced automatic fingerprinting system---which we call \ourtool{}---to effectively address the three main issues stepwise, as shown in Figure~\ref{fig:system_overview}.

\noindent\textbf{Phase-1: SD-WAN Control Traffic Classification.}
To address C1 in Section~\ref{sec:system_challenge}, adversaries first need to prune data traffic noise from control traffic traces. It is worth noting that SD-WAN control traffic displays a periodical time-series pattern along with multi-directional relationships among nodes. To extract those features, \ourtool{} first runs a coarse-grained classification as shown in Figure~\ref{fig:system_overview}. The \emph{Packet Preprocessor} module arranges traffic traces into 2-tuple (\texttt{SrcIP, DstIP}) flows. The \emph{Feature Extractor} module then generates time-series features, multi-directional features, and the session information of the given 2-tuple flows automatically. Finally, the \emph{Sequence Classifier} module solves a 2-class classification problem (i.e., SD-WAN control traffic or data traffic) based on the received features of the 2-tuple flows (see Section~\ref{subsec:classifying_sdwan}).

\noindent\textbf{Phase-2: Protocol Classification.} To address C2 in Section~\ref{sec:system_challenge}, adversaries require more fine-grained features and a distinctive deep learning model that captures the characteristics of the cluster management protocols. For this, \emph{Packet Preprocessor} module organizes the SD-WAN control traffic flows obtained in Phase-1 based on the following 5-tuple (\texttt{SrcIP, SrcPort, DstIP, DstPort, Proto}). Subsequently, the \emph{Feature Extractor} module extracts time-series and multi-directional features from the 5-tuple flows. Note that the multi-directions between all nodes are used to understand the operational characteristics of the predicted SD-WAN control traffic. Finally, the \emph{Sequence Classifier} module solves a multi-class classification problem to distinguish between the cluster management protocols using the features of given 5-tuple flows (see Section~\ref{subsec:classifying_sdwan}).

\noindent\textbf{Phase-3: Role Inference.} To address C3 in Section~\ref{sec:system_challenge}, it is necessary to discover which node is acting as the primary node and which are secondary nodes. Typically the primary nodes tend to transmit/receive more packets than their secondary nodes. Based on this, adversaries can apply the standard \emph{z-score} normalization in order to detect an outlier that is far from the average. The \emph{Role Detector} module calculates the \emph{z-score} based on the number of packets sent by each \texttt{SrcIP} address. Subsequently, it determines an inferred role for each node (e.g., primary, secondary) using \emph{z-score} thresholds. Because this procedure is simplified into a numerical measurement, it can achieve $\mathcal{O}(1)$ computational complexity. In contrast, LSTM, for example, spends redundant time for inference with $\mathcal{O}(M \times W)$, where $M$ denotes the number of LSTM layers and $W$ denotes the total number of parameters~\cite{sak2014long}.

\section{Methodology}

In this section, we present pivotal structures of \ourtool{}. Specifically, we first address the problem of how to characterize time-series and directional features. Then, we introduce a deep learning method for two different models (i.e., one for Phase-1 and the other for Phase-2) processing those features for protocol classification. Finally, we present key insight to infer the role of each node from the classified protocol flow.

\subsection{Feature Extraction}
\label{subsec:feature definition}

One of the most important tasks in \ourtool{} is feature selection. In particular, if we do not carefully characterize time-series patterns, it is likely to be a laborious task even though we develop a robust classifier model. For this, we center around the cluster management protocols' inherent traffic patterns and compare them with several control traffic to obtain insights about how traffic patterns are different between the SD-WAN control traffic and data traffic (please refer to Figure~\ref{fig:pattern_overview} in Appendix~\ref{sec:feature_comparison}). Also, we analyze multi-directional relationships, which can be observed mainly due to the SD-WAN structure. Consequently, we can prudently examine unique patterns that SD-WAN control traffic flow retains.

\noindent\textbf{Sequential Features.}
We start by considering features like \emph{the number of packets} (e.g., PPS) and \emph{packet size} (e.g., BPS) since they are commonly used when fingerprinting network traffic~\cite{schuster2017beauty}. Given that SD-WAN control traffic is delivered on layer-7, we also measure \emph{payload length} and \emph{the number of generated sessions}. Both features can represent the unique characteristics during SD-WAN operation. By analyzing these sequential features, we discover that the cluster management protocols generate packets periodically for state consistency and membership inspection. The noticeable \emph{periodic} patterns of control traffic compared to common data traffic are confirmed by the empirical results (see Figure~\ref{fig:pattern_overview}a, b, and c in Appendix~\ref{sec:feature_comparison}).

\begin{figure}[t]
    \centering
    \includegraphics[width=0.95\linewidth]{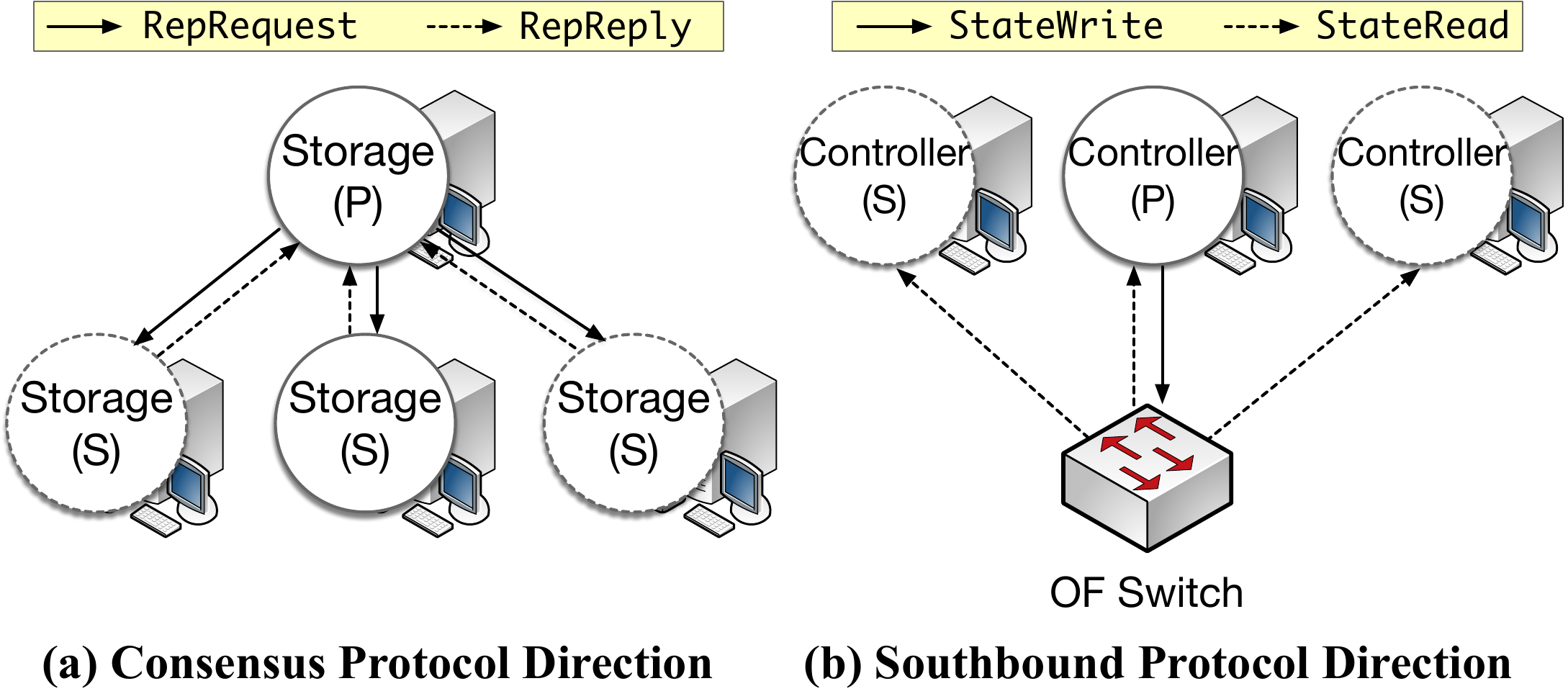}
    
    \footnotesize{$\ast$ \textbf{P}: Primary, \textbf{S}: Secondary}
    
    \caption{Examples to illustrate the insight to determine multi-directional relationships among storages-storages and controllers-switches.}
    
    \label{fig:my_label}
    \vspace{-0.15in}
\end{figure}

\noindent\textbf{Directional Features.} We observe that the SD-WAN cluster nodes show a prominent \emph{directional} pattern derived from their assigned roles. For example, the consensus protocol shows primary-centric traffic exchange patterns as shown in Figure~\ref{fig:my_label}a. Thus, it is easy to see that the primary node receives much more control packets than the secondary nodes. In Figure~\ref{fig:my_label}b, the southbound protocol shows a directional pattern between controllers and switches through the mastership operation. We can see that only the primary node has a priority to transmit write-operation to switches. By observing this pattern, adversaries can reveal which controller owns the mastership, indicating the most crucial one among controller instances. With this regard, we consider the forward (i.e., request) and the backward (i.e., reply) directions of flows transferred between multiple pairs of all nodes as \emph{multi-directional} features, which effectively track the operational aspects of SD-WAN control traffic.

\subsection{Protocol Classification}
\label{subsec:classifying_sdwan}

In this section, we describe the steps followed by \ourtool to identify which cluster management protocols are being used in an SD-WAN. We divide the classification process into two phases: (i) pruning data traffic and (ii) classifying SD-WAN cluster management protocols.

\noindent\textbf{Pruning Data Traffic.}
In order to identify SD-WAN control traffic and eliminate irrelevant (data) traffic, we collectively utilize the time-series sequence features, multi-directional features, and session information extracted by the \emph{Feature Extractor} module. As a first step, the time-series features of a data point (i.e., a single 2-tuple flow) are first split into $T$ time steps. The time-series feature vector at $t$ is constructed as follows:
$
    \mathbf{v}_t = [x^t_{bps}, \; x^t_{pps}, \; x^t_{len}],
$
where $t \in \{1, 2, ..., T\}$ denotes each time step and $x^t_{<\cdot>}$ represents the value of a given temporal feature at time $t$.
The input sequence of a given 2-tuple flow is defined as $S = [\mathbf{v}_1, \mathbf{v}_2, ..., \mathbf{v}_T]$. As Long Short-Term Memory (LSTM) has been proven as an effective recurrent neural network capable of learning long-term dependencies of sequential data~\cite{karim2017lstm}, our \emph{Sequential Classifier} module mainly consists of Bidirectional-LSTM (Bi-LSTM)~\cite{graves2013speech}, the state-of-the-art sequence processing neural network. The extracted sequence $S$ is fed into the Bi-LSTM layer, and we take the last hidden layer (i.e., $[\mathbf{h}^f_T, \mathbf{h}^b_T]$, concatenation of the last hidden state vectors of the forward and backward directions) as a sequence representation vector $\mathbf{e}_s \in \mathbb{R}^m$.

The number of created sessions is embedded into a vector $\mathbf{e}_{\sigma} = x_{\sigma} \times \mathbf{\Omega}$ using a learnable embedding vector $\mathbf{\Omega} \in \mathbb{R}^l$, where $x_\sigma$ is a generated session number created between two nodes within a specific time period.

Lastly, inspired by the multi-hot encoding~\cite{zhou2018deep}, we devise a multi-directional encoding to reflect the variety of underlying directional relationships between nodes. We choose the top-$k$ nodes which have the largest number of flows with other nodes because SD-WAN cluster nodes tend to create sessions continuously and maintain a large amount of flows. Thus, selecting $k$ number of nodes enables us to efficiently represent the flow structures of the traffic among the massive number of nodes. The \emph{top-$k$ multi-direction} of each \texttt{SrcIP} is encoded into a vector $\delta_{\texttt{SrcIP}}^{k} = [d_1, d_2, ..., d_k]$ using ternary values (i.e., $\{0, \pm 1\}$). As for $d_i$, it is defined as follows:
\begin{align*}
        d_i = 
    \begin{cases}
        +1, & \text{if a flow } \texttt{SrcIP} \rightarrow i\text{'th node exists,} \\
        -1, & \text{else if a flow } i\text{'th node} \rightarrow \texttt{SrcIP} \text{ exists,}\\
        \;\; 0, & \text{otherwise}
    \end{cases}
\end{align*}
where $i \in \{1, 2, ..., k\}$. A top-$k$ multi-directional representation vector is obtained by feeding $\delta_{\texttt{SrcIP}}$ into a dense layer $\mathit{f}$.

The sequence representation vector $\mathbf{e}_s$, the session embedding vector $\mathbf{e}_\sigma$, and the top-$k$ multi-directional representation vector are concatenated to a single vector $\mathbf{e}_p = [\mathbf{e}_s, \mathbf{e}_\sigma, \mathit{f}(\delta_{\texttt{SrcIP}})]$. It is fed into the output layer and \textit{Sigmoid} activation function, and then the model finally predicts whether the given 2-tuple sample is related to SD-WAN. For the training process, we optimize the negative log-likelihood $\mathcal{L}$ for the binary classification task (i.e., discovering if the traffic transmitted in a given flow corresponds to control traffic or not), which is defined as:
\begin{align*}
    \mathcal{L} = - \frac{1}{N} \sum^{N}_{i=1}{y_i \cdot log(\hat{y}_i) + (1 - y_i) \cdot log(1 - \hat{y}_i)},
\end{align*}
where $N$ is the total number of the samples, $y_i$ is the true label, and $\hat{y}_i$ is the predicted label of the sample $i$.

\noindent\textbf{Classifying Cluster Management Protocols.}
From the SD-WAN control traffic identified in the previous phase, we generate another deep learning model using time-series sequence and multi-directional features with the goal of distinguishing between the various cluster management protocols used within the SD-WAN. We create the sequence representation vector $\mathbf{e}_s$ consisting of time-series features for a given 5-tuple flow, which is different from the feature extraction process used in the previous phase. We apply multi-directional encoding to all SD-WAN nodes to completely track the operational characteristics of cluster management protocols, which can be formalized as $\delta_{\texttt{SrcIP}} = [d_1, d_2, ..., d_n]$, where $n$ is the number of all nodes. The multi-directional representation vector is concatenated with the sequence representation vector $\mathbf{e}_s$, then we finally obtain a vector $\mathbf{e}_p = [\mathbf{e}_s, \mathit{f}(\delta_{\texttt{SrcIP}})]$, where $\mathit{f}$ denotes a dense layer to construct the multi-direction representation vector. The final vector is fed into the output layer and activated by the \textit{Softmax} function. For a multi-class classification task, we opted to minimize the Cross-Entropy loss function defined as follows:
\begin{align*}
    \mathcal{L} = - \sum^{C}_{c=1} y_{i,c}log(\hat{p}_{i, c}),
\end{align*}
where $C$ denotes the total number of the categories, $y_{i,c}$ is $i$-th sample which belongs to category $c$, and $\hat{p}_{i,c}$ is a predictable distribution of the sample. In this context, $\hat{p}_{i,c}$ would be computed by the \textit{Softmax}.

\begin{figure}[t]
    \centering
    \subfloat[Primary-secondary flows between storages.]{
        \includegraphics[width=0.45\linewidth]{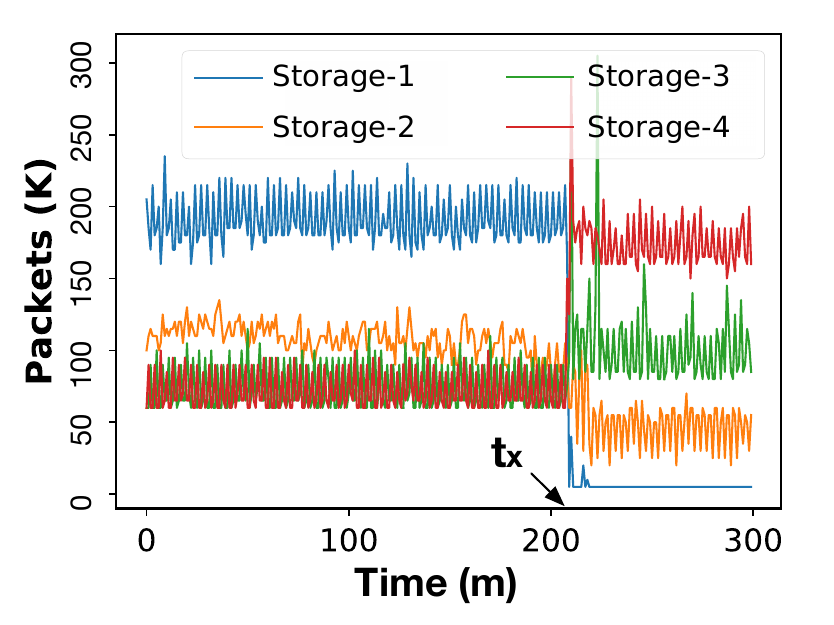}
        \label{fig:pattern_a_result}
    }
    \hspace{0.1in}
    \subfloat[Primary-secondary flows between controllers-switches.]{
        \includegraphics[width=0.45\linewidth]{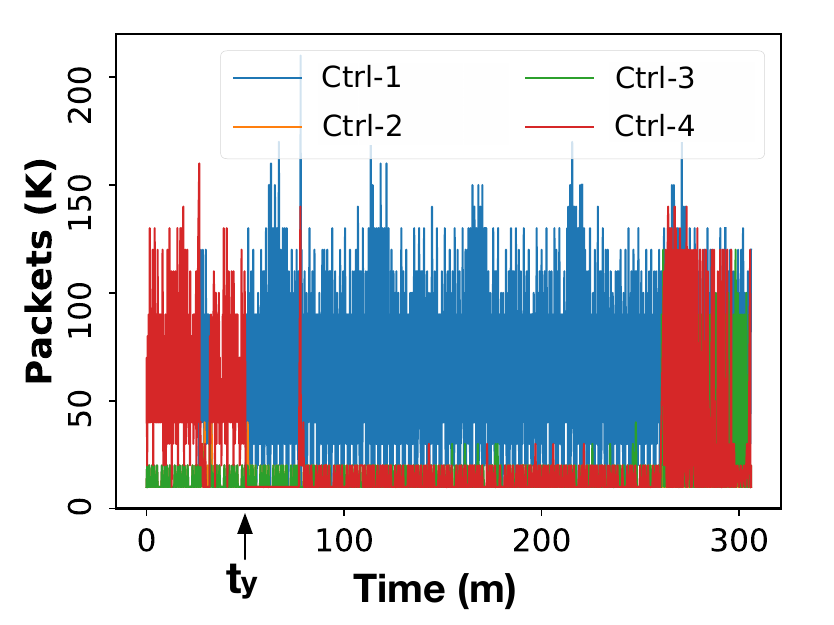}
        \label{fig:pattern_b_result}
    }
    \caption{Examples of flow patterns for different roles.}
    \vspace{-0.2in}
    \label{fig:pattern_results}
 
\end{figure}

\subsection{Role Inference}
\label{subsec:inferring_node_relations}
The roles as the primary and secondary can be distinguished through the total number of the packets they transmit. Figure~\ref{fig:pattern_a_result} illustrates the analysis of SD-WAN cluster flows concerning transmitted packets between storage nodes. \texttt{Storage-1} takes the role of a primary node at first, but then we intentionally disconnect it off (at $\mathbf{t}_x$) to observe the behavior of the secondary nodes when that happens. Specifically, we see that the remaining secondary nodes try to become the primary node for some time; finally Storage-4 is designated as the new primary node and maintains its stable connection again. Figure~\ref{fig:pattern_b_result} shows a snapshot of eavesdropped traffic flows between controllers and switch. At the beginning of the experiment, \texttt{Ctrl-4} is a primary node, so the switch only receives write-operation from the node. If we intentionally disconnect the primary node (at $\mathbf{t}_y$), one of the secondary nodes, \texttt{Ctrl-1}, will be appointed as a new primary node and send messages to modify states until the prior primary node, \texttt{Ctrl-4}, is restored. The primary nodes transmit at least twice as much data as the secondary nodes, which is always applicable, regardless of SD-WAN cluster size and accidental disconnection.

\noindent\textbf{Z-score Normalization.} Based on this insight, we utilize the \emph{z-score} normalization to define a role for each node to improve the efficiency of system flow. The utilization of \emph{z-score} normalization has proved a significant effect on classifying with prominent pattern features~\cite{singh2015bikesh}. Here, the distinct values of each node, which is organized from the classified cluster management protocol traffic, are normalized based on a mean and standard deviation. The \emph{z-score} normalization value $z$ is defined as $z = (x - \mu) / \sigma$, where $x$ is the distinct values, $\mu$ is the mean of our data, and $\sigma$ is the standard deviation. In this regard, we first arrange the total number of transmitted packets for each \texttt{SrcIP}. The total number of the packets of each \texttt{SrcIP} can be the distinct value $x$ with which its mean value can be easily calculated. The calculated score enables us to infer the roles by defining a threshold ($\theta_z$) in a heuristic manner.

%% file: 4_eval.tex
\section{Evaluation}
\label{sec:evaluation}

In this section, we evaluate the three phases mentioned above considering various environments to demonstrate their practicability. We first evaluate precision, recall, and F1-score to classify SD-WAN control and data traffic. Then, we show the results of classifying cluster management protocols. Additionally, we demonstrate that our classification method effectively represents the sequential and multi-directional features in comparison with several existing works' methods, such as a hybrid model of traditional time-series modeling algorithm (i.e., auto-regressive integrated moving average) and a different deep learning model. Additionally, we demonstrate the robustness of our classification model by testing the accuracy while adopting several defense systems. After that, we calculate \emph{z-score} based on the total number of the packets for each refined \texttt{SrcIP} to infer one of the roles as a primary (i.e., leader, master) or a secondary (i.e., follower, slave) role for each node. Finally, we achieve control-plane topology and protocol dependencies. 

\begin{figure}[t]
    \centering
    \includegraphics[width=0.85\linewidth]{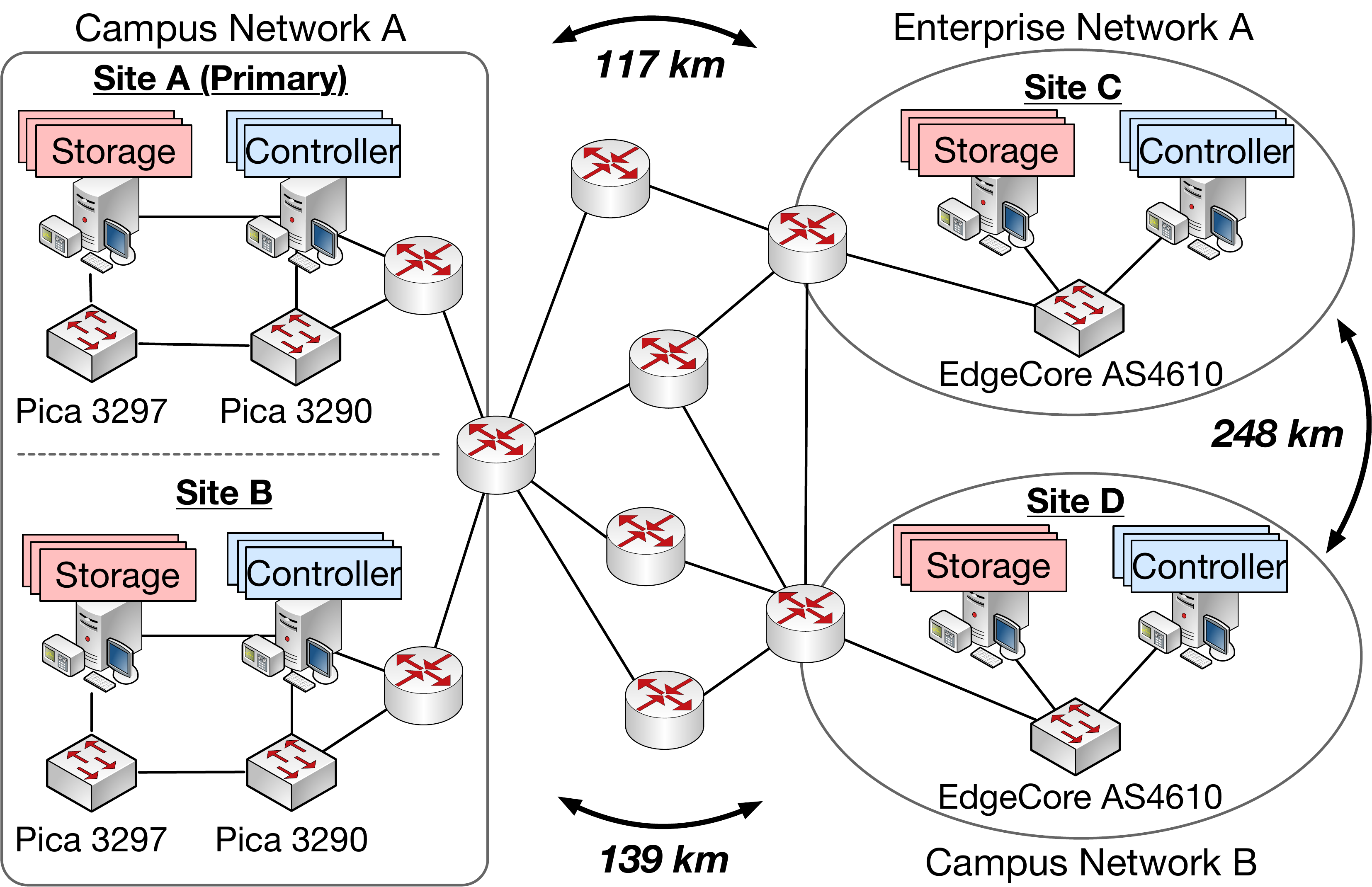}
    
    \caption{Our SD-WAN testbed. It consists of three physically distant sites, and each site is composed of two physical machines and (an) SDN switch(es).}
    \label{fig:testbed_overview}
    \vspace{-0.2in}
\end{figure}

\subsection{Experimental Environment}
\label{sec:experimental_environment}
We construct a realistic SD-WAN testbed with four different sites comprising three campus networks and one enterprise network, with an average distance of 170 kilometers (see Figure~\ref{fig:testbed_overview}), using multiple machines equipped with Intel Xeon Silver 4210R CPUs (10 cores) and 64 GB RAM. All these machines run distributed controllers for each site. Each site uses one or two commercial SDN-enabled switches (e.g., Pica 3297, EdgeCore AS4610), which are controlled by multiple storages and controllers. It is worth noting that our methodology can be used in any targeted SD-WAN environment, but for generality, we utilize Atomix v3.1.5~\cite{atomix_web} as storage nodes and ONOS v2.2.0~\cite{onos_web} as controller nodes, which are the most popular open source tools. ONOS controller deploys Raft~\cite{ongaro2014search} and SWIM ~\cite{das2002swim} protocols for the consensus and membership protocols, respectively. Also, ONOS controller uses OpenFlow (OF)~\cite{openflow_1_3}, a de-facto standard in SDN, as the southbound protocol. Note that the Raft and OF protocol distinguish primary-secondary roles into \emph{leader-follower} storages and \emph{master-slave} controllers, respectively. 
As for adversaries, we envision that they will be able to create their own SD-WAN environments (as we did in our experiments) with multiple configurations to obtain labeled data and create necessary pre-trained models for fingerprinting SD-WAN (see Appendix~\ref{sec:labeled_samples}).

We use NVIDIA GeForce RTX 3090 GPUs for deep learning models and select classification methods with LSTM (50 seconds window size). We conduct experiments with 10, 50, 100, 200, 400 of dimensions of the LSTM hidden layers, $1 \times 10^{-4}$, $5 \times 10^{-4}$, $1 \times 10^{-3}$, $2 \times 10^{-3}$, $4 \times 10^{-3}$ of learning rates, and 32, 64, 128, 256, 512 of batch sizes. Then, we choose the best conditions: 200 of dimensions, $1 \times 10^{-4}$ of learning rate, and 256 of batch size (list it with more details, as well as of different baseline models, in Appendix~\ref{sec:hyperparam}). In the training and evaluation process, we randomly select 70\% samples as a train set and 30\% samples as a test set. We employ 7-fold cross validation on the train set to train a model without any biases that the dataset may contain.

\input{table_1_phase_1}

\subsection{Dataset Description}
To our best knowledge, public SDN control traffic has not been released. Therefore, it is significant to collect control traffic data as reliable as we can.

\noindent\textbf{SD-WAN Control Traffic.} 
\label{sec:dataset_visibility}
We collect a variety size of storages and controllers' traffic deploying many containers in each machine by utilizing the container management tool Docker~\cite{docker_web} because the sizes of SD-WAN clusters are numerous depending on demanding services. We use Atomix~\cite{atomix_web} to produce storage control traffic and ONOS~\cite{onos_web} to generate controller traffic. In the meantime, we execute six ONOS applications (e.g., \texttt{Reactive Forwarding}, \texttt{OpenFlow Driver}, \texttt{Access Control}, \texttt{Control Message Stats Provider}, \texttt{DHCP Service}, \texttt{Mastership Load Balancer}) on our testbed to reflect various data-plane events. Thus, we consider as many environments as possible extensively.

\noindent\textbf{Data Traffic.} We utilize CAIDA backbone traffic~\cite{caida_2014,caida_2015,caida_2016} to provide a realistic network environment in our testbed. CAIDA passive trace dataset includes traces collected from high-speed monitors on a real backbone link. To evaluate our fingerprinting method fairly, we mix CAIDA traces with the SD-WAN control traffic. Also, we play streaming videos from a secured VPN machine and Tor browser, and we utilize an email server and online-chatting platform (e.g., Skype) to improve the reality of our testbed. Furthermore, to evaluate the effectiveness of our system fairly, it is necessary to include the traffic that exhibits periodical patterns with SD-WAN traffic, such as the ones generated from commercial distributed systems. Therefore, we measure and store \texttt{Hyperledger}~\cite{hyperledger_web} control traffic which is developed for a suite of stable frameworks for enterprise-grading blockchain deployments. \texttt{Hyperledger} is currently deploying the Raft protocol for the consensus algorithm, so that it can be proper control traffic to evaluate our system's effectiveness. Also, we collect \texttt{ZooKeeper}~\cite{zookeeper_web} control traffic, which is supporting distributed synchronization and group services by connecting all clients using a mechanism of membership inspection.

\noindent\textbf{Dataset Collection Methodology.} We collect 53,880,288 data traffic packets containing various services/events and 3,343,093 SD-WAN control packets from our testbed which comprises 4 nodes to 20 nodes. Also, \emph{Feature Extractor} module generates 81,487 data traffic flow dataset and 3,096 SD-WAN control traffic flow dataset. The SD-WAN control traffic flows including 4 nodes to 20 nodes in our testbed are used to emulate a realistic environment of the SD-WAN cluster having various sizes. Moreover, we create 5-tuple based dataset reflecting multi-directional relationships for the train and test dataset, which are labeled with three classes (i.e., Raft, SWIM, OpenFlow). Here, we perform our experiment with three test cases (i.e., $\mathbf{T1}$, $\mathbf{T2}_p$, $\mathbf{T2}_s$) to study the feasibility of the attacks under different settings where the adversaries obtain different amount of data. Specifically, we consider the $\mathbf{T1}$ test case where network adversaries can eavesdrop traffic on multiple sites. Also, we consider the $\mathbf{T2}_p$ and $\mathbf{T2}_s$ test cases where on-site adversaries reside in the site with primary node and the site with secondary node, respectively.
\vspace{-0.05in}

\subsection{Accuracy of SD-WAN Control Traffic Classification} 
Table~\ref{tab:phase1_result} describes how effectively SD-WAN control traffic is distinguished from data traffic. We use an auto-regressive integrated moving average (ARIMA) algorithm with decision tree classifier~\cite{sutoyo2020hybrid, xu2021novel} and a convolutional neural network (CNN)~\cite{lecun1998gradient,wang2017end,lotfollahi2020deep} classifier as baselines (through following the method of existing works) to show that our classifier can reflect time-series and multi-directional patterns effectively.

Overall, our LSTM-based classifier can identify both data traffic and SD-WAN control traffic with high accuracy. As for classifying SD-WAN control traffic in the $\mathbf{T1}$ test case, the F1-score is 96.08\% for the average of executing 100 times, manifesting the practicality of our model. As for the $\mathbf{T2}_s$ test case, despite maintaining high accuracy outcomes on classifying data traffic, identifying SD-WAN control traffic has relatively lower accuracy on precision, recall, and F1-score. Since the $\mathbf{T2}_s$ test case captures partial traffic, it has constraints to reflect the time-series and directional patterns generated by unseen nodes sufficiently in comparison with the $\mathbf{T1}$ and $\mathbf{T2}_p$ test cases. However, the average F1-score is 91.75\% when executed 100 times, so it still demonstrates the consistency of our model.
The ARIMA-based classifier is difficult to reflect complex traffic patterns, so that it maintains overall lower F1-score than that of our LSTM-based classifier when classifying SD-WAN control traffic. In the case of the CNN-based classifier, classifying data traffic exhibits overall similar accuracy compared to our LSTM-based classifier in the three test cases. However, the F1-scores of classifying SD-WAN control traffic are quite lower compared with the LSTM-based classifier. This is because the explicit time-series and multi-directional patterns of SD-WAN control traffic cannot be represented by CNN that effectively reflects local and position-invariant features~\cite{lopez2017network}.

\input{table_2_phase_2}

\vspace{-0.1665in}

\subsection{Accuracy of Protocol Classification}
In Phase-2, we show the effectiveness of classifying the cluster management protocols. Table \ref{tab:phase2_result} illustrates the performance between the three classifiers under different settings, and among them, our LSTM-based classifier achieves overall higher accuracy than those derived from the ARIMA-based and the CNN-based classifiers.

In the case of our classifier targeting the $\mathbf{T1}$ test case, F1-score for each of the Raft, SWIM, and OpenFlow protocol is 80.73\%, 81.92\%, and 90.78\%, respectively. It manifests a high performance because it can reflect the context of time-series patterns and flow directions sufficiently in SD-WAN. Also, the F1-scores of the OpenFlow protocol are higher than those of the Raft and SWIM protocols. This is because the OpenFlow protocol maintains a long-term connection compared with the SWIM protocol and exhibits more explicit flow directions of master/slave nodes comparing with the Raft protocol.

The $\mathbf{T2}_p$ test case cannot observe various flow directions of the control traffic as sufficient as the $\mathbf{T1}$ test case. Nevertheless, our system successfully considers the unique pattern of flow directions generated by the cluster management protocols, so that it can maintain satisfactory accuracy for classifying the Raft protocol. The SWIM protocol classification in the $\mathbf{T2}_p$ test case exhibits a relatively lower F1-score compared with the $\mathbf{T1}$ test case as we cannot capture sufficient control traffic among other follower-to-follower traffic connections in the $\mathbf{T2}_p$ test case.

Consider the case where the traffic becomes sparser, like $\mathbf{T2}_s$ test case, the ARIMA-based classifier exhibits slightly higher accuracy only in the SWIM protocol classification than those of the CNN-based and the LSTM-based classifiers. This is because both classifiers cannot reflect the directional patterns of the SWIM protocol, which creates sessions ephemerally and disappears in the short term. However, all the rest of the protocols in our LSTM-based classifier exhibit much higher F1-score. For example, classifying the Raft protocol shows approximately 19\%p and 6\%p higher F1-scores than that of the ARIMA-based classifier and the CNN-based classifier, respectively. Regarding OpenFlow protocol classification, the F1-scores are approximately 29\%p and 18\%p higher than that of the ARIMA-based classifier and the CNN-based classifier, respectively. Thus, our LSTM-based classifier is the most powerful classifier to reflect the characteristics of flow directions as well as the time-series patterns generated by the cluster management protocols.

\vspace{-0.1in}

\subsection{Robustness against Defense Systems}
Network operators may adopt several defense systems to hinder adversaries from learning unique patterns of network traffic. To demonstrate the robustness of our deep learning-based fingerprinting under the defense systems, we consider the case where network operators impose random noise and perturbation mechanism~\cite{wang2014effective, gong2020zero, wang2017walkie, rastogi2010differentially} on the outgoing traffic. We first add random noise, 10\% extent from the root mean square (RMS) of the training inputs. In addition, we impose Fourier Perturbation Algorithm (FPA)~\cite{rastogi2010differentially} on the input traffic, which is a largely adopted method to protect sequence data. FPA is formalized by $\mathbf{FPA} = \mathbf{IDFT}(\textbf{LPA}(\textbf{DFT}(\text{\textit{input}}), \lambda))$, where $\mathbf{DFT}$ and $\mathbf{IDFT}$ denote the Discrete Fourier Transform (DFT) and the Inverse DFT, respectively, and $\mathbf{LPA}$ denotes the Laplace Perturbation Algorithm. The $\lambda$ is set to 30\% of the RMS of the training inputs. 

\begin{figure*}[t]
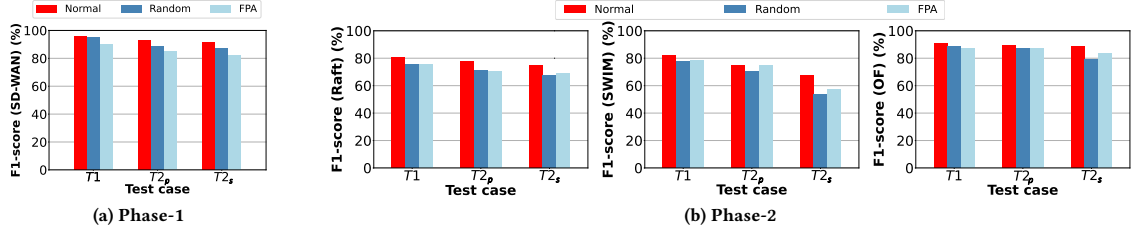

    \centering
    \subfloat[Phase-1]{
       \includegraphics[width=0.193\linewidth]{figures/noise_phase_1_v3.pdf}
     }
     \vspace{-0.05in}
     \hspace{0.5cm}
     \subfloat[Phase-2]{
        \includegraphics[width=0.6\linewidth]{figures/noise_phase_2_v3.pdf}
    }
    \vspace{-0.05in}
    \caption{The accuracy of classifying SD-WAN traffic (Phase-1) and cluster management protocols (Phase-2) while several defense mechanisms are adopted.} 
    \footnotesize{$\ast$ \textbf{Random}: Random noise, \textbf{FPA}: Fourier Perturbation Algorithm}
    \label{fig:noise}
    \vspace{-0.1in}
\end{figure*}

In Figure~\ref{fig:noise}, our model for Phase-1 exhibits robust performance, such as 93.14\% and 91.75\% of F1-scores in the $\mathbf{T1}$ test case, against random noise and FPA, respectively. Our model for Phase-2 also well classifies cluster management protocols while keeping only a 4.2\%p difference compared to our normal model (without defense systems) in the $\mathbf{T1}$ test case. The reason is that the directional features can effectively reflect the unique operational characteristics derived from SD-WAN. Therefore, we achieve reasonable performance while having sufficient directional information of the cluster management protocols' operations (e.g., in $\mathbf{T1}$ and $\mathbf{T2}_p$ test cases) even though the sequence data is perturbed by the defense systems.

\begin{figure}[t]
    \centering
    \subfloat[Consensus/membership control traffic distribution as function of \emph{z-score}.]{
       \includegraphics[trim=0.1cm 0 0 0, width=0.85\linewidth]{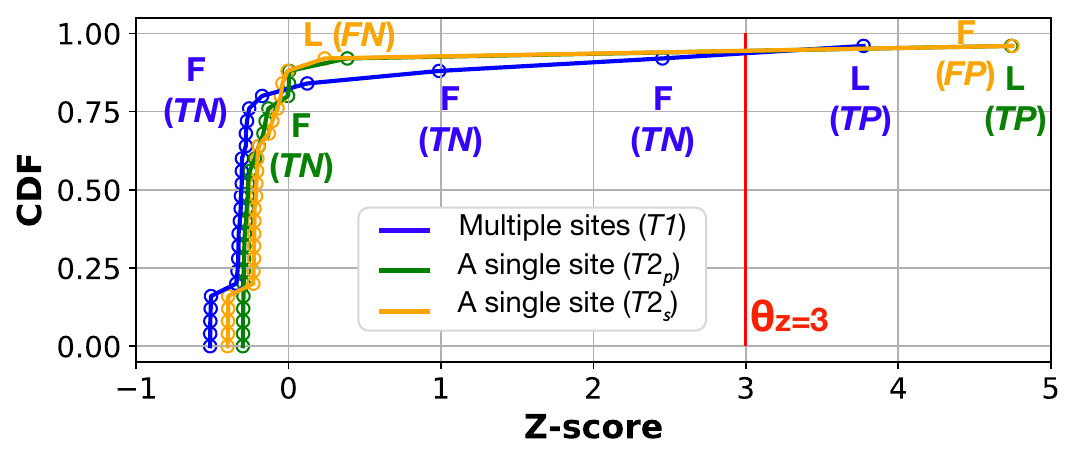}
     }
     \vspace{-0.1in}
     \subfloat[Southbound control traffic distribution as a function of \emph{z-score}.]{
        \includegraphics[trim=0.1cm 0 0 0, width=0.85\linewidth]{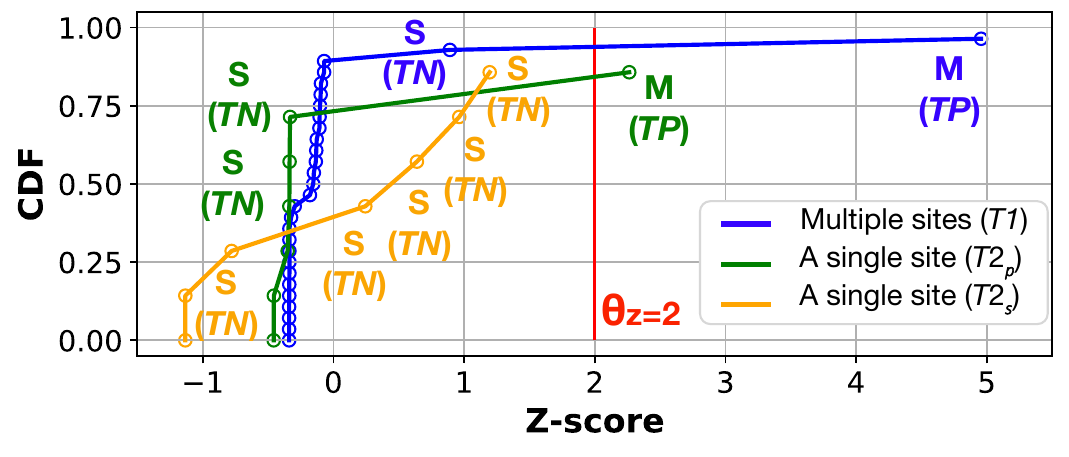}
     }
    
    \footnotesize{$\ast$ \textbf{L}: Leader, \textbf{F}: Follower, \textbf{M}: Master, \textbf{S}: Slave, \\
    \textbf{TP}: True Positive, \textbf{FP}: False Positive, \textbf{FN}: False Negative, \textbf{TN}: True Negative}
    \vspace{-0.1in}
    \caption{Measured control traffic distributions observed in three test cases. The marked dots denote \emph{z-score} of storage/controller nodes. The labels indicate inference results when the threshold $\theta_z$=3 and $\theta_z$=2 (the red lines) for detecting storage/controller roles, respectively.}
    \label{fig:ew_zscore}
    \vspace{-0.1in}
\end{figure}

\subsection{Effectiveness of Role Inference}

We now evaluate how \ourtool{} correctly infers a role for each node using the \emph{z-score} normalization based on the results derived from our LSTM-based classifier that showed the highest accuracy in the previous phases. In our analysis, the primary (i.e., leader, master) nodes are more likely to be located above the threshold ($\theta_z$) in contrast with the secondary (i.e., follower, slave) nodes, which is computed by True Positive Rate (TPR) and False Positive Rate (FPR) comprehensively. Also, each execution time for calculating \emph{z-scores} and detecting roles takes less than 2 seconds.

In Figure~\ref{fig:ew_zscore}a, it shows that \emph{z-scores} for inferring leader/follower roles are 3.77, 4.74, and 4.75 for the $\mathbf{T1}$, $\mathbf{T2}_p$, and $\mathbf{T2}_s$ test case, respectively. Based on the aforementioned \emph{z-scores}, role inferences from the $\mathbf{T1}$ and $\mathbf{T2}_p$ test cases are placed at our predictable domain. This is because the control traffic is almost included in the $\mathbf{T1}$ and $\mathbf{T2}_p$ test cases due to their wide coverage as the leader node transmits replication state messages constantly to its follower nodes. However, the control traffic generated from a follower node can be observed only by other follower nodes, so it cannot reflect the sufficient context of flow directions from the $\mathbf{T2}_s$ test case.

The \emph{z-scores} for inferring master/slave roles are 4.95, 2.26, and 1.20 for the $\mathbf{T1}$, $\mathbf{T2}_p$, and $\mathbf{T2}_s$ test cases, respectively (see Figure~\ref{fig:ew_zscore}b). Since the specific controller node identified in the traffic flow generated from the $\mathbf{T1}$ test case can achieve the mastership on several switches, master/slave relationships inferred from the $\mathbf{T1}$ test case show a noticeable aspect considering the entire network context. Similarly, we also figure out that the specific controller node identified in traffic flows generated from the $\mathbf{T2}_p$ test case can have the mastership on multiple switches. Hence, it is straightforward to distinguish between master and slave in the $\mathbf{T1}$ and $\mathbf{T2}_p$ test cases based on a reasonable threshold. However, the traffic flow generated from the $\mathbf{T2}_s$ test case only reflects the network context partially, so we can only define the roles of slaves.

\subsection{Similarity of Control-Plane Topology}
\label{ss:similarity}
We finally evaluate how similarly adversaries can reconstruct control-plane topology and protocol dependencies. To quantitatively evaluate this, we model them as a graph abstraction:

\noindent\textbf{Graph.} A graph is denoted by $G=(V,E)$, where $V$ denotes a set of protocol processes (e.g., Raft, SWIM, OpenFlow) running upon cluster nodes, and $E$ denotes a set of protocol dependencies between the two nodes. In Figure~\ref{fig:protocol_dependency}, the edge $Con_{Raft (P)} \leftrightarrow Con_{Raft (S)}$ shows a Raft protocol dependency between the leader (primary) and follower (secondary) nodes. The graph shows topological information of the cluster nodes by analyzing how vertices/edges are composed in the graph.

To compare the similarity, we use Graph Edit Distance (GED)~\cite{gao2010survey} metric that shows the number of graph operations (e.g., node insertion/deletion, edge insertion/deletion). This metric can represent how much cost we require for transforming the inferred topology into the real one from a graph structure perspective. To define the similarity, suppose that there are two different graphs $G_1$ and $G_2$. The topology similarity between the two graphs can be defined as:
$
    Sim(G_1, G_2) = 1 - \frac{GED(G_1, G_2)}{|V_1| + |E_1| + |V_2| + |E_2|},
$
where the $GED$ denotes the graph edit distances, $V_{i\in \{1, 2\}}$ is a set of protocol processes belonging to the inferred cluster nodes specified with their roles, and $E_i$ is a set of the protocol dependencies between nodes of a graph $G_i$.
To compute GEDs, we utilize GEDEVO~\cite{ibragimov2013gedevo}, which is for rapidly calculating GEDs of such large-sized graphs.

\begin{figure}[t]
    \centering
    \includegraphics[width=0.92\linewidth]{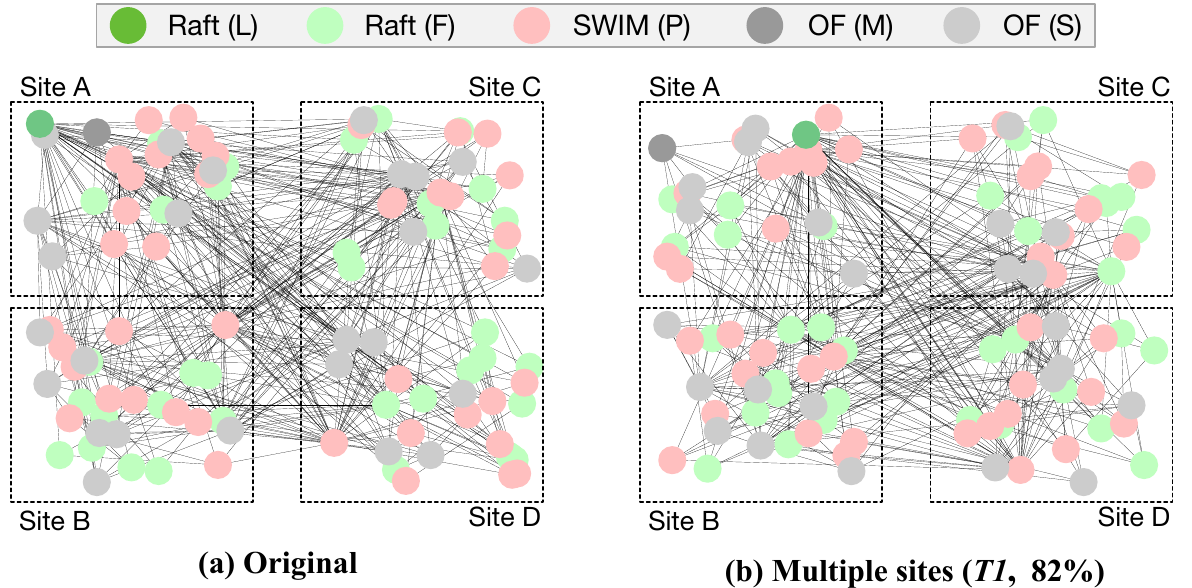}
    
    \footnotesize{$\ast$ \textbf{($n$\%)} denotes similarity percentage for reconstructing to \textbf{(a) Original} SD-WAN testbed as shown in Figure~\ref{fig:testbed_overview}.}
    \caption{Snapshots of the inferred control-plane topology and protocol dependencies.}
    \label{fig:topology_snapshot}
    \vspace{-0.05in}
\end{figure}

We configure a total of 20 storage/controller nodes, which are constructed with five storage/controller nodes for each site to simulate the inference scenario in our testbed. As shown in Figure~\ref{fig:topology_snapshot}a, the leader (i.e., green) and master (i.e., dark grey) nodes are located at Site A originally. When targeting the $\mathbf{T1}$ test case in Figure~\ref{fig:topology_snapshot}b, the similarity is 82\% and shows the highest accuracy in comparison with other test cases because the control traffic is captured by the $\mathbf{T1}$ test case mostly, so that it can consider the time-series patterns and the context of the flow directions abundantly compared with other test cases (please refer to Appendix~\ref{sec:reconstructed_topo}).

%% file: table_1_phase_1.tex
\begin{table*}[t]
\centering
\footnotesize
\caption{Classifying SD-WAN control traffic and data traffic.}
\begin{tabular}{ c c c c c c c c c c c }

\toprule
\multirow{2}{*}{\textbf{Test Case}} & \multirow{2}{*}{\textbf{Traffic Type}} & \multicolumn{3}{c}{\textbf{ARIMA-based Classifier}~\cite{sutoyo2020hybrid, xu2021novel}} & \multicolumn{3}{c}{\textbf{CNN-based Classifier}~\cite{wang2017end,lotfollahi2020deep}} & \multicolumn{3}{c}{\textbf{LSTM-based Classifier (Ours)}}   \\

\cmidrule{3-11} 
& & Precision & Recall & F1-score & Precision & Recall & F1-score & Precision & Recall & F1-score \\
\midrule

\multicolumn{1}{c}{\multirow{1}{*}{\emph{Multiple sites}}} & Data & 88.32\% & 94.73\% & 90.91\% & 99.12\% & 99.40\% & 99.18\% & 99.70\% & 99.78\% & 99.32\% \\

($\mathbf{T1}$) & SD-WAN & 92.28\% & 79.44\% & 85.30\% & 90.62\% & 86.83\% & 88.79\% & 96.73\% & 95.57\% & \textbf{96.08}\%   \\

\midrule

\multicolumn{1}{c}{\multirow{1}{*}{\emph{A single site}}} & Data & 84.54\% & 96.02\% & 89.87\% & 98.88\% & 99.62\% & 99.28\% & 99.89\% & 99.88\% & 99.82\%  \\

($\mathbf{T2}_p$) & SD-WAN & 91.37\% & 72.59\% & 80.90\% & 90.76\% & 76.39\% & 82.75\% & 93.04\% & 93.74\% & \textbf{93.14}\%   \\

\midrule
\multicolumn{1}{c}{\multirow{1}{*}{\emph{A single site}}} & Data & 80.17\% & 94.13\% & 86.03\% & 95.58\% & 99.16\% & 97.37\% & 99.82\% & 99.84\% & 99.79\% \\

($\mathbf{T2}_s$) & SD-WAN & 88.76\% & 69.24\% & 77.79\% & 91.46\% & 89.29\% & 89.35\% & 93.68\% & 86.03\% & \textbf{91.75}\% \\


\bottomrule

\end{tabular}
\label{tab:phase1_result}
\vspace{-0.1in}
\end{table*}

%% file: table_2_phase_2.tex
\begin{table*}[t]
\centering
\footnotesize
\caption{Classifying cluster management protocols.}
\begin{tabular}{ c c c c c c c c c c c }

\toprule

\multirow{2}{*}{\textbf{Test Case}} & \multirow{2}{*}{\textbf{Traffic Type}} & \multicolumn{3}{c}{\textbf{ARIMA-based Classifier}~\cite{sutoyo2020hybrid, xu2021novel}} & \multicolumn{3}{c}{\textbf{CNN-based Classifier}~\cite{wang2017end,lotfollahi2020deep}} & \multicolumn{3}{c}{\textbf{LSTM-based Classifier (Ours)}}  \\

\cmidrule{3-11}

& & Precision & Recall & F1-score & Precision & Recall & F1-score & Precision & Recall & F1-score  \\
\midrule

\multicolumn{1}{c}{\multirow{3}{*}{\begin{tabular}{c} \emph{Multiple sites} \\ ($\mathbf{T1}$) \end{tabular}}} & Raft & 63.18\% & 55.73\% & 59.66\% & 68.55\% & 81.60\% & 74.47\% & 81.67\% & 78.39\% & \textbf{80.73}\% \\

& SWIM & 76.95\% & 79.24\% & 78.37\% & 69.86\% & 75.16\% & 72.50\% & 78.28\% & 85.18\% & \textbf{81.92}\% \\
& OpenFlow & 79.10\% & 50.21\% & 61.18\% & 90.13\% & 60.64\% & 72.46\% & 86.04\% & 95.57\% & \textbf{90.78}\% \\

\midrule

\multicolumn{1}{c}{\multirow{3}{*}{\begin{tabular}{c} \emph{A single site} \\ ($\mathbf{T2}_p$) \end{tabular}}} & Raft & 56.98\% & 40.19\% & 46.27\% & 66.95\% & 76.83\% & 71.60\% & 78.92\% & 76.15\% & \textbf{77.95}\% \\

& SWIM & 74.52\% & 59.84\% & 65.36\% & 67.57\% & 68.81\% & 68.47\% & 76.01\% & 72.24\% & \textbf{74.68}\% \\
& OpenFlow & 68.77\% & 62.13\% & 64.54\% & 85.07\% & 60.54\% & 71.01\% & 84.21\% & 95.19\% & \textbf{89.13}\%  \\

\midrule
\multicolumn{1}{c}{\multirow{3}{*}{\begin{tabular}{c} \emph{A single site} \\ ($\mathbf{T2}_s$) \end{tabular}}} & Raft & 36.12\% & 30.74\% & 33.87\% & 71.73\% & 68.15\% & 69.99\% & 78.26\% & 70.52\% & \textbf{74.43}\% \\

& SWIM & 69.92\% & 68.21\% & \textbf{69.04}\% & 73.26\% & 61.09\% & 66.25\% & 65.91\% & 68.02\% & 67.11\% \\
& OpenFlow & 77.81\% & 67.24\% & 71.16\% & 89.18\% & 57.45\% & 69.98\% & 86.92\% & 90.25\% & \textbf{88.43}\% \\



\bottomrule

\end{tabular}
\label{tab:phase2_result}
\vspace{-0.10in}
\end{table*}

%% file: 5_use_case.tex
\section{\ourtool{} Use Case}
\label{sec:use_cases}

\begin{figure}[t]
    \centering
    \includegraphics[width=.92\linewidth]{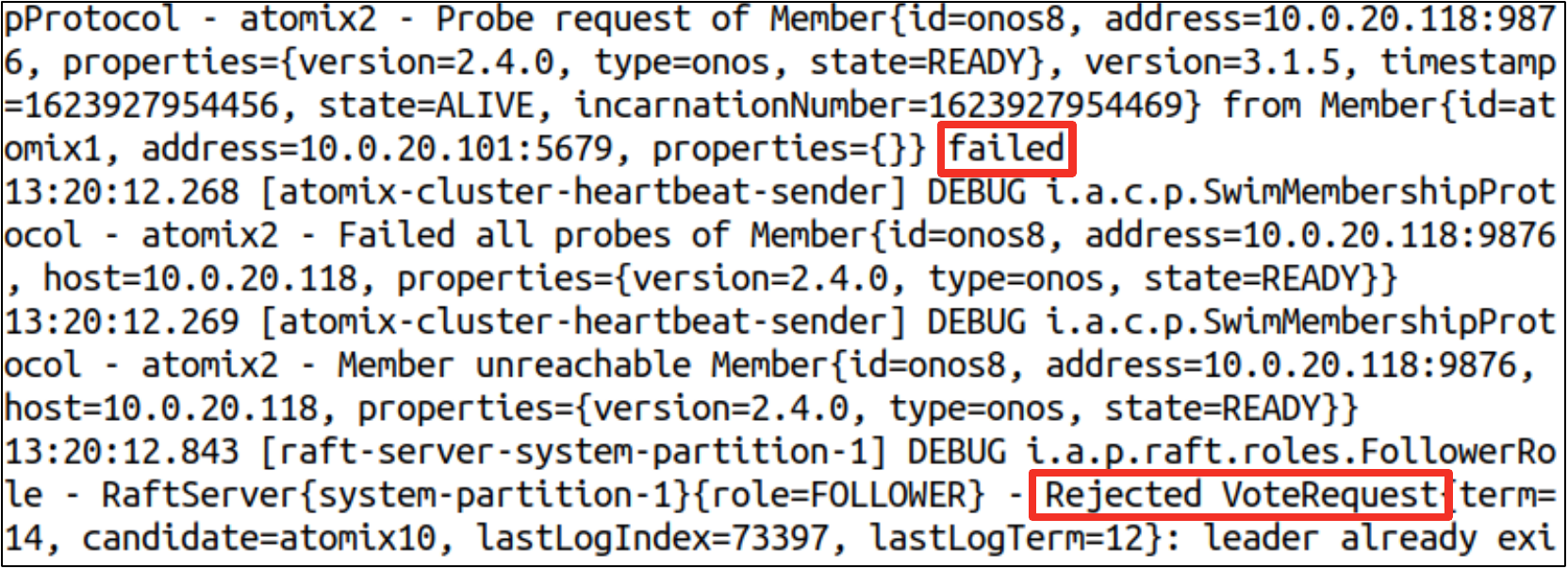}
    \caption{Cluster management communication failures observed in Atomix logs.}
    \label{fig:attack_result}
    \vspace{-0.1in}
\end{figure}

Based on the achieved confidential information, adversaries can now specify the critical path through which important cluster management protocol messages pass the most. To demonstrate the feasibility of such vulnerabilities, we present a stealthy DDoS attack scenario in which adversaries flood a shared path that is a part of in-band\footnote{The prior study~\cite{cao2019crosspath} showed that adversaries can locate in-band paths where control traffic and data traffic share the same bandwidth. Specifically, adversaries can measure timing differences of RTTs by sending end-to-end probing---that trigger flow rule installation---to a target path. If different RTTs are observed, it means that the probing interferes control traffic; thus, we can infer that the path contains in-band links.} cluster management channel between cluster nodes. Because mounting a large volume of traffic in real-world WAN can raise major ethical issues, we establish a private SD-WAN testbed (see Figure~\ref{fig:local_testbed} in Appendix~\ref{sec:interruption}).

Now, adversaries can conduct an advanced disrupting approach in a stealthy manner. After they specify the target services or protocols currently deployed and define a role for each node, adversaries may impose a low volume of traffic from a number of adversarial bots (eventually becomes an enormous traffic) on the shared link which is near from the \emph{leader} node. Subsequently, it can cause vote failures across the entire SD-WAN cluster as shown in Figure~\ref{fig:attack_result}.

%% file: 6_related_work.tex
\section{Related Work}

It is worth noting that many prior studies have demonstrated the severity if unique traffic patterns are exposed to adversaries. Although the traffic is under SSL/TLS encryption, the threats can be applicable on both legacy network and SDN-based network.

\noindent\textbf{Encrypted Traffic Analysis.} 
Roei et al.~\cite{schuster2017beauty} demonstrated that adversaries can perceive burst patterns of video streams to identify which video files are currently being accessed by users. Sun et al. \cite{sun2019automated} proposed a data-driven approach for passive fingerprinting of IoT devices based on the classification of encrypted SSL/TLS flows generated from several IoT devices. Apthorpe et al. \cite{apthorpe2017smart} showed that analyzing the network traffic rate of smart home devices can reveal sensitive user interactions even when the traffic is encrypted. Some studies \cite{cai2012touching, wang2014effective, al2016adaptive} de-anonymized end-hosts from encrypted traffic such as Tor. These prior researches motivate our approach and show that even though SDN control channels are secured by SSL/TLS, they can be in danger if their unique patters are exposed.

\noindent\textbf{Fingerprinting in SDN.} 
Various security experts have actively focused on fingerprinting methodology on SDN due to its unique structural characteristics. For example, timing-based end-to-end reconnaissance can inform adversaries about SDN internal system state, the behavior of the controller, host communication pattern, and detection of target flows~\cite{kloti2013openflow, liu2017flow, sonchack2016timing,shin2013attacking}. Even though previous studies could be practical in a single type of protocol messages (e.g., OpenFlow), they still have constraints on fingerprinting control traffic that includes mixed-protocol packets in a single connection, such as cluster management protocols. To address the problem, a deep learning-based approach has emerged in recent years. Cao et al.~\cite{cao2019fingerprinting} introduced a possible attack scenario that analyzes OpenFlow traffic and infers running applications on a target SDN controller using deep learning models. The deep learning-based approach could classify encrypted network control packets and infringe on network privacy in a single controller environment by applying a time-series pattern of OpenFlow traffic. However, the approach is not applicable to a multi-controller environment because the approach could not consider multi-directional control traffic generated from intricate interactions between various nodes (e.g., controllers, storages) in the cluster.

%% file: 7_conclusion.tex
\section{Conclusion}
We devise a novel tool, \ourtool{}, which enables infringing on SD-WAN confidential information. We accumulate enormous data traffic and actual SD-WAN control traffic from our realistic SD-WAN testbed. \ourtool{}, then, learns unique time-series patterns for fingerprinting SD-WAN control traffic automatically and multi-directional relationships for protocol classification. We test the feasibility of our system under different settings (i.e., traffic collection from a single site or multiple sites) and under an environment where several defense systems are adopted to compare its contrast effectiveness. Thus, we demonstrate that adversaries can infer control-plane topology and protocol dependencies with reasonable accuracy. Finally, we present critical attack vectors that can be abused practically in SD-WAN. We believe that our study provides security insights to SD-WAN operators by foreseeing and testing the potential risks of the system in advance.

\begin{acks}
This work was supported by the National Research Foundation of Korea (NRF) grant funded by the Korea government (MSIT) (No. RS-2022-00166401). The research leading to these results have received funding from the European Union’s Horizon 2020 research and innovation programme under grant agreements No 871793 (Accordion), No 101016509 (Charity) and No 101070473 (FLUIDOS).
\end{acks}

%% file: 8_appendix.tex
\appendix

\section{Obtaining Labeled Samples}
\label{sec:labeled_samples}
Before we infer confidential information, we first need a pre-trained model, which is trained offline, without knowing any information about the targeted SD-WAN cluster. Fortunately, the number of controllers and cluster management protocols being used is relatively small, so we can potentially consider them all in the offline model. To this end, we monitor ongoing data traffic and SD-WAN control traffic in the self-made environment. The constructed environment and configuration settings can be varied over a large range of cluster sizes (e.g., using container management tools~\cite{docker_web} or establishing the real world testbed) depending on what we aim to achieve. These various environment settings with different combinations of data traffic enable our offline model to be more robust and to consider labeled samples from as many controllers/protocols as possible. 

\section{Long-term Distinguishable Features}
\label{sec:feature_comparison}
\begin{figure}[h]
    \centering
    \normalsize
    \includegraphics[trim=0.2cm 0 0.5cm 0,clip,width=\linewidth]{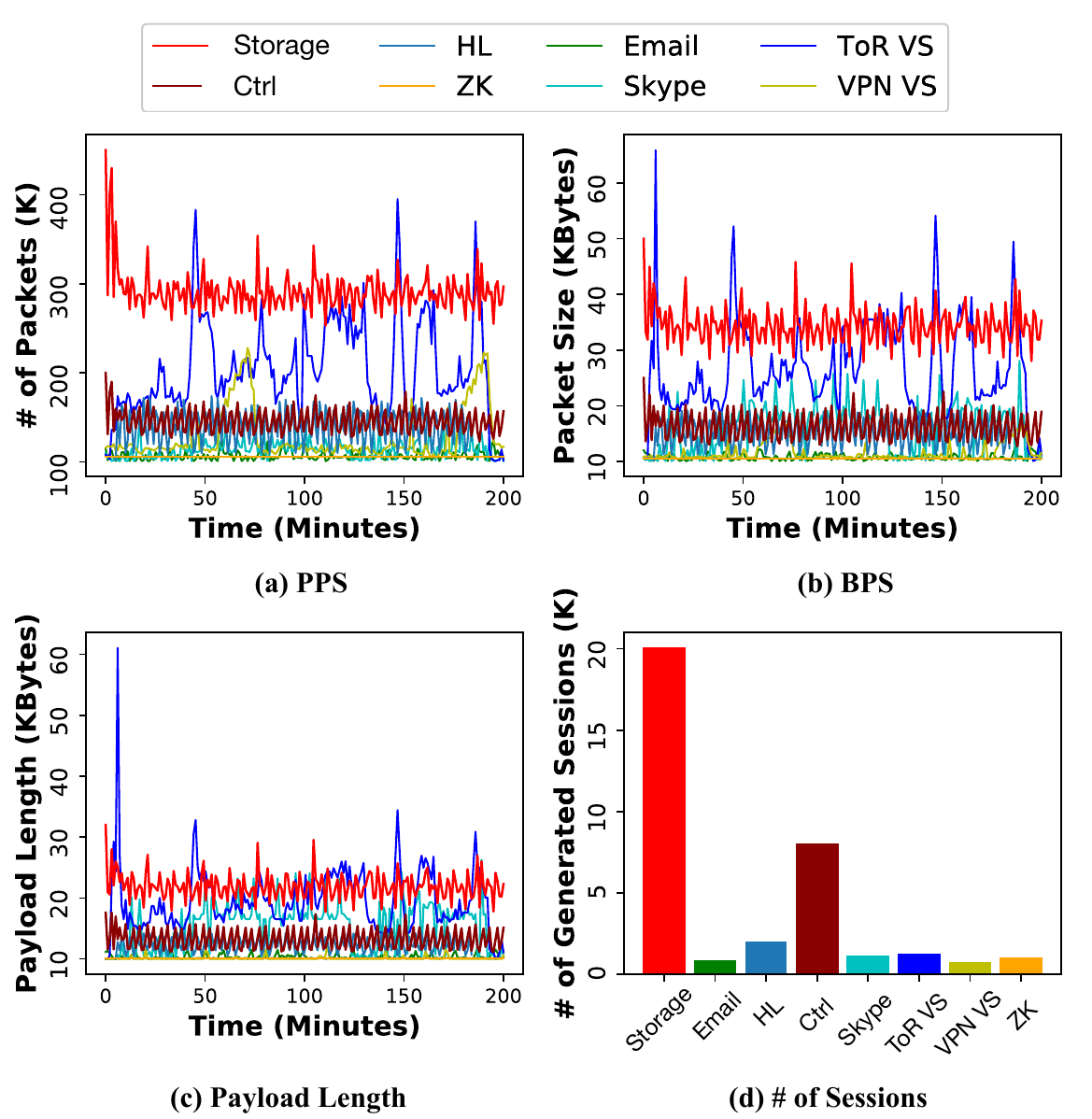}
    \footnotesize{$\ast$ \textbf{HL}: Hyperledger, \textbf{ZK}: Zookeeper, \textbf{VS}: Video Streaming}
    \caption{Displaying distinguishable traffic patterns for four attributes: \textit{\# of packets}, \textit{packet length}, \textit{payload length}, \textit{\# of generated sessions}.}
    \label{fig:pattern_overview}
\end{figure}

In Figure~\ref{fig:pattern_overview}, whereas other distributed systems (e.g., \texttt{ZooKeeper}, \texttt{Hyperledger}) exhibit stable patterns, we discover that the cluster management protocols (i.e., \texttt{Storage}, \texttt{Ctrl}) are more likely to display noticeable \emph{periodic} patterns. The reason is that the consensus protocol and membership protocol generate packets periodically for its state consistency and membership inspection. Storage nodes and controller nodes create sessions constantly whenever they exchange messages; thus, the cluster management protocols tend to be positioned quantitatively higher than other consensus algorithms. In addition, the rest of the data traffic (e.g., Email, Skype, ToR Video Streaming, VPN Video Streaming) illustrate totally different aspects.
Specifically, they show a fluctuated pattern comparing with the aforementioned consensus protocols. Therefore, our targeted-features can maintain quite stable in the long term. Also, the cluster management control traffic creates sessions constantly and disappears shortly after exchanging messages (see Figure~\ref{fig:pattern_overview}d). Thus, embedding generated session numbers for each control traffic flow can make our classifier models more robust.

\section{Comparison of Reconstructed Control-Plane Topology}
\label{sec:reconstructed_topo}

\begin{figure}[!ht]
    \centering
    \includegraphics[width=\linewidth]{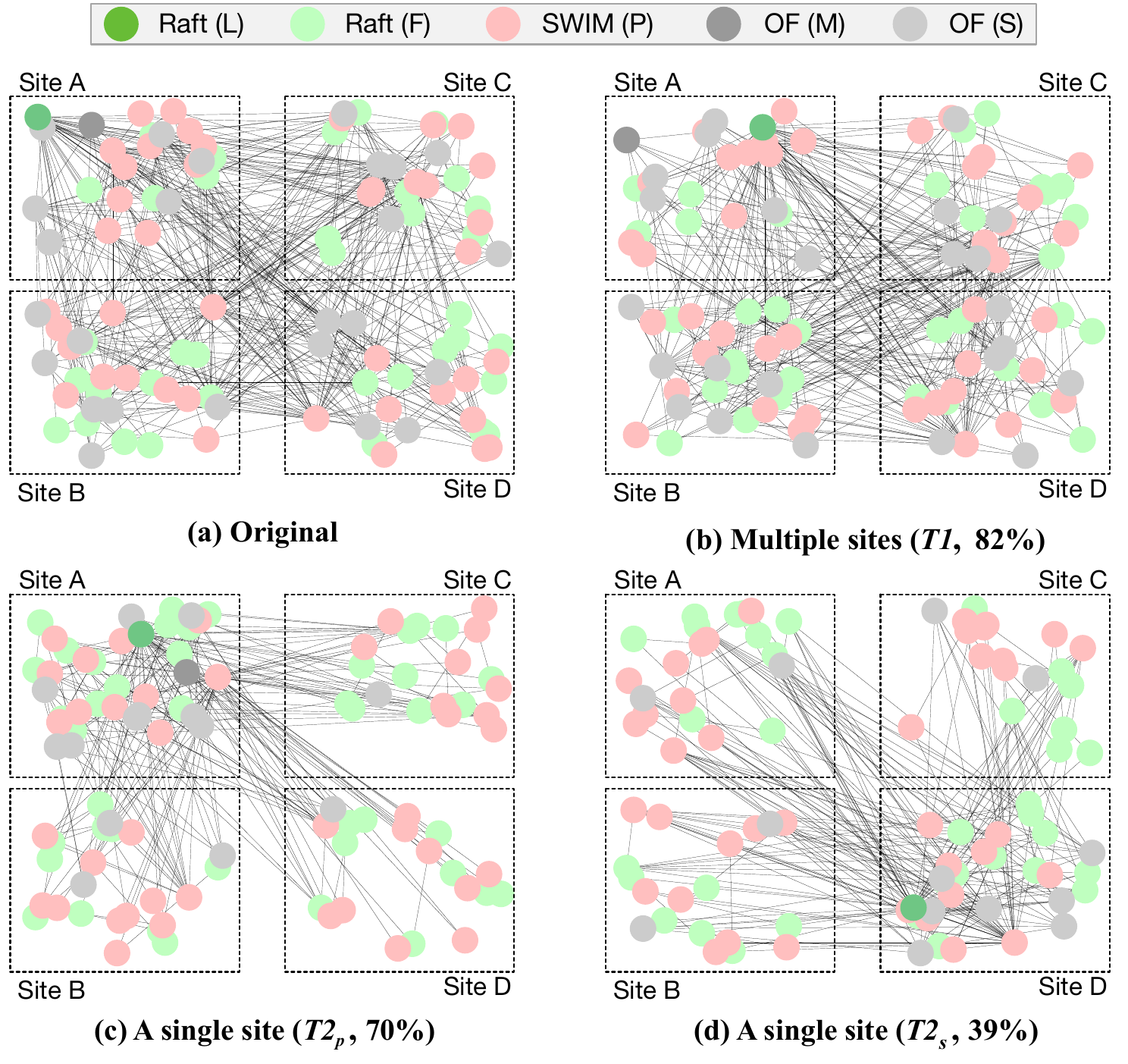}
    \footnotesize{$\ast$ \textbf{($n$\%)} denotes similarity percentage for reconstructing to \textbf{(a) Original} SD-WAN testbed as shown in Figure~\ref{fig:testbed_overview}}
    \caption{Snapshots of the inferred control-plane topology and protocol dependencies.}
    \label{fig:topology_snapshot_apdx}
\end{figure}

As we mentioned in Section~\ref{ss:similarity}, \textbf{T1} test case could reconstruct SD-WAN control-plane topology with high accuracy. Here, we exhibit snapshots of the inferred control-plane topology and protocol dependencies captured in different sites. As for the $\mathbf{T2}_p$ test case in Figure~\ref{fig:topology_snapshot_apdx}c, the similarity is 70\%, relatively lower accuracy comparing with the $\mathbf{T1}$ test case. This is because control traffic of the Raft protocol collected from the $\mathbf{T2}_p$ test case exchanges messages between the leader and the follower nodes in the entire network. Also, the control traffic collected from the $\mathbf{T2}_p$ test case cannot include SWIM protocol traffic generated among other follower-to-follower traffic connections. In the end, the similarity evaluated from the $\mathbf{T2}_s$ test case has the lowest accuracy (i.e., 39\%) because it cannot infer a proper role for each node in Phase-3, so it misidentifies one of the nodes in its local site (i.e., Site D) as a leader.

\section{Control channel Interruption}
\label{sec:interruption}

\begin{figure}[t]
    \centering
    \includegraphics[width=0.95\linewidth]{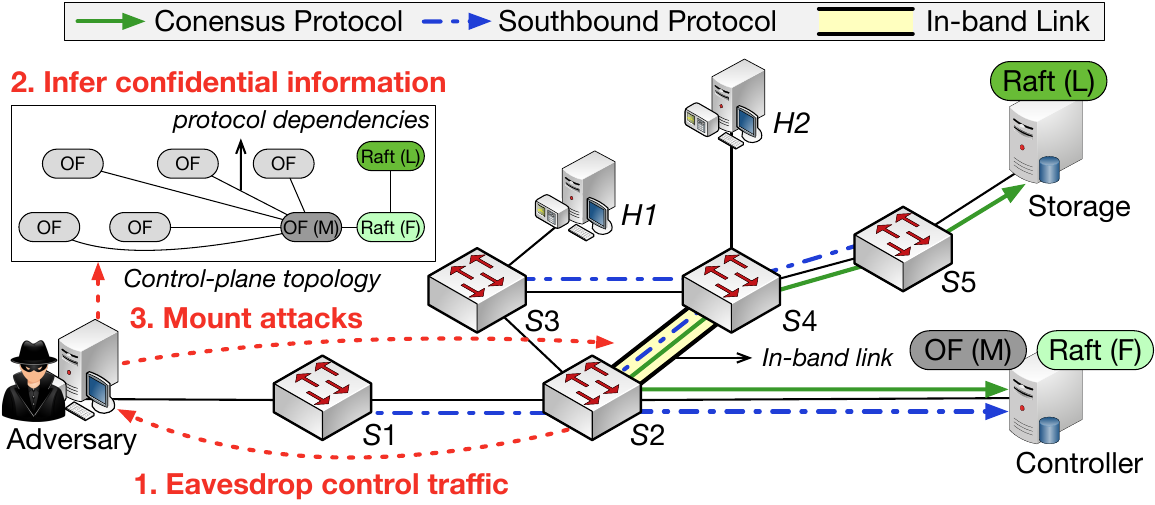}
    \caption{An example that shows how an adversary disrupts cluster communication on an in-band link $S2$-$S4$ between the Storage (leader) and Controller (follower). Note that the in-band link carries cluster management control traffic and data traffic at the same time.}
    \label{fig:local_testbed}
\end{figure}

After fingerprinting the control-plane topology and protocol dependencies, adversaries can conduct harmful attacks that disrupt the operation of an SD-WAN. Due to the ethical issue when we generate a large volume of traffic in a real-world WAN environment, we establish our private SD-WAN testbed as shown in Figure~\ref{fig:local_testbed}.

Our private testbed consists of five Pica 3297 switches, a commercial SDN switch, and five Intel Xeon Silver 4210R CPUs by allocating three CPUs for adversaries' machines and two CPUs for running SDN storages and controllers. We deploy various containers to configure a number of adversarial bots and support a multi-controller environment in each server, by utilizing the container management tool Docker~\cite{docker_web}. Subsequently, we generate a low volume of traffic for each single link but impose an enormous volume of traffic on the shared path. To impose the volumetric traffic and demonstrate prominent performance difference effectively, all of the links are connected with 10 GbE interfaces.

\begin{figure}[t]
    \centering
    \includegraphics[width=.95\linewidth]{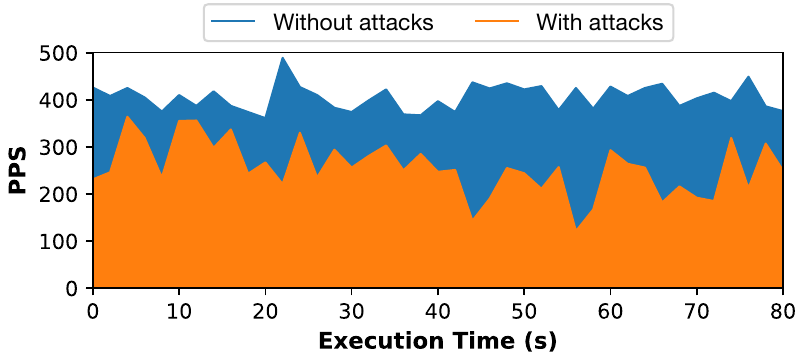}
    \caption{Measured throughput (PPS) of the cluster management protocol traffic for two cases; \emph{(i) without attacks} and \emph{(ii) with attacks}.}
    \label{fig:attack_plot}
\end{figure}

As we mentioned above, our attack could cause vote failures. As a result of the failures, it shows that the attack causes a maximum of 71\% and an average of 37\% performance degradation, as shown in Figure~\ref{fig:attack_plot}.

\section{Mitigation}

In this section, we present a possible countermeasure to prevent or make it more difficult for adversaries to gain any insights about the topology and the underlying protocols of an SD-WAN from the (encrypted) communication patterns. We mainly discuss the ones based on sending SD-WAN traffic across multiple network routes (in different ASes).

\noindent\textbf{Multi-path Routing.} At first glance, one way to mitigate the identified privacy issues would be to preclude adversaries from collecting enough traffic such that they are unable to create an accurate fingerprint of the SD-WAN. The experiments we conducted in Section~\ref{sec:evaluation} show that, while the fingerprinting tasks are very accurate if adversaries have access to exchanged packets sufficiently, these tasks become more difficult when adversaries only have access to a small portion of the packets transmitted. From this observation and inspired by new Internet architectures like SCION~\cite{10.1145/3085591}, it becomes apparent that multi-path routing can be an effective mechanism to raise the bar to adversaries who aim to fingerprint the SD-WAN. With multi-path routing, the SD-WAN East-West traffic exchanged between any two sites would be sent across a sufficiently large number of network routes leveraging already-established agreements between ASes. The main advantage of this defense is that it does not require changing the nodes’ behaviors or the underlying protocols to reduce the amount of data exposed to adversaries. However, this mechanism provides some protection under the assumption that the adversaries can only gather traffic at one or a few locations.

\noindent\textbf{Obfuscation-based Methods.} Unlike the previous method, which does not tackle the root of the problem but rather aims to hinder the adversaries from obtaining sufficient data for the fingerprinting, obfuscation-based mechanisms focus on eliminating or considerably reducing the information revealed to adversaries through the (encrypted) communication patterns (e.g.,~\cite{meier2022ditto}). These mechanisms generally rely on delaying network packets, padding packets, or creating new dummy packets with the goal of perturbing the traffic features the adversaries can leverage for fingerprinting. While such methods can reduce information leakage to a large extent, they also come with important limitations. For example, due to the importance of the traffic exchanged between multiple SD-WAN sites (for synchronizing internal states and checking the membership), delaying network packets may not be a viable option since this could negatively affect network performance, availability, and security. Similarly, both padding techniques and the addition of dummy packets entail a significant increase in communication cost, hampering their adoption in practice. Furthermore, most obfuscation-based methods require significant modifications to existing protocols (or need to deploy modern network devices with specific capabilities), which could be a laborious task and hinder practicality in the real world. This suggests that lightweight obfuscation mechanisms, which can resolve key limitations of existing methods (e.g., degrading of network performance and wasting bandwidth), may be further developed to target \ourtool{} fingerprinting attack.

\section{Hyperparameters}
\label{sec:hyperparam}

We automatically search the optimal order $(p,d,q)$ for each time-series data (where $p, d$, and $q$ denote the auto-regressive order, the degree of differencing, and the moving average order, respectively, with $p$ and $q$ values starting from 1 to 3) and select the ARIMA model that exhibits the lowest Akaike’s information Criterion (AIC) value. Then, we use a decision tree classifier with default settings of criterion and maximum depth (based on the fact that nodes can be expanded until all leaves are pure). As we mentioned in Section~\ref{sec:experimental_environment}, we select the best LSTM hyperparameter settings: 200 of dimensions, $1 \times 10^{-4}$ of learning rate, and 256 of batch size. As for the CNN model, we conduct experiments with $1 \times 10^{-4}$, $5 \times 10^{-4}$, $1 \times 10^{-3}$, $2 \times 10^{-3}$, $4 \times 10^{-3}$ of learning rates, 32, 64, 128, 256, 512 of batch sizes, 10, 50, 100, 200, 400 of the size of hidden layers, and 1, 2, 3 of kernel size. Then, we choose the best CNN hyperparameter settings: $1 \times 10^{-3}$ of learning rate, 256 of batch size, 100 of the size of hidden layers, and 2 of kernel size (see Table~\ref{tab:hyperparameter}).

\input{table_3_hyperparameter}

\section{Discussion}
\label{sec:discussion}

\noindent\textbf{Performance While Deploying Defense Systems.} We discover that the accuracy of SWIM protocol in $\mathbf{T2}_s$ test case (see Figure~\ref{fig:noise}) can be slightly lower than the other test cases because the experiment is based on data collected in a location where no primary nodes reside. In addition, the traffic of SWIM protocol exhibits (i) short-term connection properties and (ii) weak directional patterns during its operation; thus, the pattern of SWIM traffic can be easily masked by defense systems. However, the overall performance of our classifier (experimented on other test cases) against defense systems exhibits reasonable accuracy.

\noindent\textbf{Automatic Control-Plane Environment Generation.}
We recognize that the generated number of control traffic is different depending on SD-WAN environments. Even though a deep learning model is known to be efficient at training time-series patterns, its effectiveness can be decided by how different environments are provided. Therefore, we need to consider as various circumstances as possible. For the better deployment of the experiment, we will develop a new framework for our future work to support as various SD-WAN environments as possible in an automated manner, such as several frameworks that generate different SDN assessment environments~\cite{lee2017delta,jero2017beads}.

%% file: table_3_hyperparameter.tex
\begin{table}[t]

\centering
\footnotesize

\caption{Hyperparameter settings of different models.}
\begin{tabular}{ c c c } 

\toprule

\multirow{2}{*}{\textbf{Hyperparameter}} & 
\multicolumn{2}{c}{\textbf{DNN Models}} \\

\cmidrule{2-3} 
& CNN & LSTM \\

\midrule
\multicolumn{1}{c}{\multirow{1}{*}{Optimizer}}      & Adam          & Adam          \\ \midrule
\multicolumn{1}{c}{\multirow{1}{*}{Learning rate}}  &  $1 \times 10^{-3}$    &  $1 \times 10^{-4}$  \\ \midrule
\multicolumn{1}{c}{\multirow{1}{*}{Batch size}}     &   256  & 256 \\ \midrule
\multicolumn{1}{c}{\multirow{1}{*}{Hidden size}}    & 100    & 200   \\ \midrule
\multicolumn{1}{c}{\multirow{1}{*}{Kernels size (CNN)}}   & 2   & -- \\ \midrule

\end{tabular}
\label{tab:hyperparameter}
\end{table}